\documentclass[apj,usenatbib,twocolumn]{aastex6}
\usepackage{float}
 \usepackage{xspace}
\usepackage{threeparttable}
\usepackage{nicefrac}
\usepackage{afterpage}
\usepackage{natbib}
\usepackage{url}

\newcommand{\angstrom}{\textup{\AA}}
\def\kms{{\rm km~s$^{-1}$}\xspace}
\def\Lsun{{\rm L$_{\odot}$}\xspace}
\def\Rsun{{\rm R$_{\odot}$}\xspace}
\def\Msun{{\rm M$_{\odot}$}\xspace}

\begin{document} 

   \title{Common Envelope ejection for a Luminous Red Nova in M101}

   \author{N. Blagorodnova \altaffilmark{1,3},
           R.~Kotak \altaffilmark{2},
           J.~Polshaw \altaffilmark{2},
           M.~M.~Kasliwal \altaffilmark{1},
           Y.~Cao \altaffilmark{1},
           A.~M.~Cody \altaffilmark{7},
           G.~B.~Doran \altaffilmark{9},
           N.~Elias-Rosa \altaffilmark{8},
           M.~Fraser \altaffilmark{3},
			C.~Fremling \altaffilmark{4},
            C.~Gonzalez-Fernandez \altaffilmark{3},
            J.~Harmanen \altaffilmark{11},
            J.~Jencson \altaffilmark{1},
			E.~Kankare \altaffilmark{2},            
            R.-P.~Kudritzki \altaffilmark{12},
			S.~R.~Kulkarni \altaffilmark{1},
            E.~Magnier \altaffilmark{12},
            I.~Manulis \altaffilmark{5},
            F.~J.~Masci \altaffilmark{1},
            S.~Mattila \altaffilmark{3,11},
            P.~Nugent \altaffilmark{6},
            P.~Ochner \altaffilmark{8},
            A.~Pastorello \altaffilmark{8},
            T.~Reynolds  \altaffilmark{11,13},
            K.~Smith \altaffilmark{2},
            J.~Sollerman \altaffilmark{4},
            F.~Taddia \altaffilmark{4},
            G.~Terreran \altaffilmark{2,8},
            L.~Tomasella \altaffilmark{8},
            M.~Turatto \altaffilmark{8},
            P.~M.~Vreeswijk \altaffilmark{5},             
            P.~Wozniak \altaffilmark{10},
            S.~Zaggia  \altaffilmark{8} }


   \altaffiltext{1}{
   Cahill Center for Astrophysics, California Institute of Technology, Pasadena, CA 91125, USA }
    \altaffiltext{2}{Astrophysics Research Centre, School of Mathematics and Physics, Queen's University Belfast, Belfast BT7 1NN, UK }
    \altaffiltext{3}{Institute of Astronomy, University of Cambridge, Madingley Road, CB3 0HA, Cambridge, UK }
     \altaffiltext{4}{ Department of Astronomy, The Oskar Klein Center, Stockholm University, AlbaNova, 10691 Stockholm, Sweden }
     \altaffiltext{5}{ Department of Particle Physics and Astrophysics, Weizmann Institute of Science, Rehovot 7610001, Israel }
     \altaffiltext{6}{ Lawrence Berkeley National Laboratory, Berkeley, California 94720, USA }
     \altaffiltext{7}{ Spitzer Science Center, California Institute of Technology, 1200 East California Boulevard, Pasadena, Ca 91125, USA }
    \altaffiltext{8}{ INAF-Osservatorio Astronomico di Padova, Vicolo dell Osservatorio 5, I-35122 Padova, Italy }
     \altaffiltext{9}{ Jet Propulsion Laboratory, California Institute of Technology, Pasadena, CA 91125, USA  }
    \altaffiltext{10}{ Los Alamos National Laboratory, MS-D466, Los Alamos, NM 87545, USA }
    \altaffiltext{11}{  Tuorla Observatory, Department of Physics and Astronomy, University of Turku, V\"ais\"al\"antie 20, FI-21500 Piikki\"o, Finland }
    \altaffiltext{12}{ Institute for Astronomy, University of Hawaii, 2680 Woodlawn Drive, Honolulu, HI 96822, USA}
    \altaffiltext{13}{ Nordic Optical Telescope, Apartado 474, E-38700 Santa Cruz de La Palma, Spain }

   \date{Received XX XX XXXX; accepted XX XX, XXXX}

  \begin{abstract}
We present the results of optical, near-infrared, and mid-infrared observations of M101 OT2015-1 (PSN J14021678+5426205), a luminous red transient in the Pinwheel galaxy (M101), spanning a total of 16 years. The lightcurve showed two distinct peaks with absolute magnitudes $M_r\leq-12.4$ and $M_r \simeq-12$, on 2014 November 11 and 2015 February 17, respectively. The spectral energy distributions during the second maximum show a cool outburst temperature of $\approx$3700 K and low expansion velocities ($\approx-$300 \kms) for the H I, Ca II, Ba II and K I lines. From archival data spanning 15 to 8 years before the outburst, we find a single source consistent with the optically discovered transient which we attribute to being the progenitor; it has properties consistent with being an F-type yellow supergiant with $L$~$\sim$~8.7~$\times\ 10^4$ \Lsun, $T_{\rm{eff}}\approx$7000~K and an estimated mass of $\rm{M1}= 18\pm 1$ \Msun. This star has likely just finished the H burning phase in the core, started expanding, and is now crossing the Hertzsprung gap. Based on the combination of observed properties, we argue that the progenitor is a binary system, with the more evolved system overfilling the Roche lobe. Comparison with binary evolution models suggests that the outburst was an extremely rare phenomenon, likely associated with the ejection of the common envelope. The initial mass of the binary progenitor system fills the gap between the merger candidates V838 Mon (5$-$10 \Msun) and NGC~4490-OT~(30~\Msun). 
\end{abstract}



\section{Introduction}

The discovery of an unusually bright and red nova in M31 (M31 RV) in September 1988 \citep{Rich1989}, triggered the attention of astronomers towards an uncommon type of object. Its peak absolute magnitude, $M_{\rm V} = -9.95$, was brighter than a regular nova ( $M_{\rm V}$ =$-6$ to $-8$), but fainter than a Supernova ($M_{\rm V}<-14$ mag). The surprisingly cool temperature, similar to an M0 type super-giant, and high ejected mass, placed the object into a potentially different category from known cataclysmic variables eruptions, triggering the need for further theoretical exploration. Since then, transient surveys and discoveries led by amateurs contributed to further populate this luminosity ``gap'' between classical novae and supernovae \citep[SNe;][]{Kasliwal2011iptf}. As of today, the observational diversity of such intermediate luminosity events on long timescales ($>$20 days) encompasses three main categories: (1) SN impostors, due to eruptions in massive stars such as luminous blue variables (LBV), (2) intermediate luminosity optical (red) transients (ILOT/ILRT), explained as terminal faint explosions and (3) luminous red novae (LRNe), which are potential stellar mergers.

Luminous non-terminal outbursts of massive stars may sometimes mimic the observational signature of a SN. Consequently, this class of events was named as ``SN impostors''. Among these, eruptions of LBVs are known to produce intermediate luminosity transients 
\citep{Humphreys1994}, such as Eta Carinae and P Cygni. These classical examples generally inhabit the upper part of the Hertzsprung-Russell (HR) diagram, having bolometric magnitudes brighter than $M_{\rm{Bol}}=-9.5$ mag, in the super-giant region.
Generally, LBV progenitors exhibit giant eruptions with visual changes $>$2 mag, but they also show non-periodic variability consistent with the behaviour of known LBVs in the LMC: R127 and S Doradus \citep[][and references therein]{Wolf1989,Walborn2008}. As a consequence, the progenitor stars are generally living in a dusty environment, caused by previous episodes of mass ejections. The non-terminal eruptions of SN 2009ip \citep{Pastorello2013,Mauerhan2013,Fraser2013} and UGC 2773 OT2009-1 \citep{Foley2011,Smith2016b} are examples of LBVs in their cool eruptive phase.

ILRT, such as SN 2008S \citep{Prieto2008,Botticella2009,Thompson2009}, NGC 300 2009OT-1 \citep{Bond2009,Smith2011} and iPTF10fqs \citep{Kasliwal2011iptf} also inhabit the luminous part of the ``gap'' transient family \citep{Kasliwal2011}. Such events have been interpreted as faint terminal explosions associated to dusty progenitors \citep{Prieto2008,Prieto2009,Kochanek2011}. The electron-capture SNe scenario has been suggested as a possible mechanism \citep{Botticella2009}. Late time observations reveal the complete disappearance of their progenitors, suggesting their outburst to be a terminal activity \citep{Adams2016}. However, NGC 300-OT has also been interpreted as due to accretion on the secondary \cite{Kashi2010}. A survey of massive stars in M33 revealed that the rate of SN 2008S and the NGC 300-OT-like transient events is of the order of $\sim20$\% of the CCSN rate in star-forming galaxies in the local Universe (D$_{\rm{L}} \lesssim 10$ Mpc) \citep{Thompson2009}. However, the fraction of massive stars with colours similar to the progenitors of these transients is only $\lesssim 10^{-4}$. \cite{Khan2010} showed that similar stars are as rare as one per galaxy. The direct implication is that the heavy dust environment phase is a very short transition phase for many massive stars during their final $10^4$ years.

Violent binary interactions in binary systems (including stellar mergers) were suggested as the plausible scenario to explain the nature of the outbursts of LRNe \citep{Iben1992,Soker2003,Tylenda2011,Ivanova2013Sci}. Nova Scorpii 2008 (V1309 Sco) currently provides the most compelling evidence for a merger scenario in our own Galaxy, as the exponential period decay of the progenitor system could be witnessed from observations spanning several years before the outburst \citep{Mason2010,Tylenda2011,Tylenda2013,Nandez2014}. V833 Mon, at 6.1$\pm$ 0.6 kpc \citep{Sparks2008} is another remarkable example of a low mass stellar merger candidate \citep{Soker2003}, including a spectacular light echo revealed by observations with the Hubble Space Telescope \citep{Bond2003}. Some extragalactic examples of discoveries consistent with the merger scenario are M85-OT2006OT-1\footnote{Although M85 2006OT-1 is observationally similar to other LRNe, its nature is more controversial: \cite{Kulkarni2007} (see also \cite{Ofek2008}, \cite{Rau2008}) supported the idea of a low-to-moderate mass merger, while \cite{Pastorello2007} favored the weak core-collapse SN explosion scenario.}, the luminous red nova in M31, reported in \cite{Kurtenkov2015} and \cite{Williams2015}, and the massive stellar merger NGC 4490 2011OT-1 \citep{Smith2016}. Pre-explosion photometry of the progenitor systems has allowed to estimate the mass and evolutionary stage of several progenitor systems. To date, the literature reports a wide range of cases, from 1.5 $\pm$ 0.5 \Msun for V1309 Sco to 20$-$30 \Msun for NGC 4490 2011OT-1 \citep{Smith2016}. In agreement with the progenitor mass function, the estimated observed Galactic rate of such events is one every few years ($\sim$ 3 yr) for low luminosity events ($M_V \geq -4$) and one every 10$-$30 yr for intermediate luminosity ($-7 \leq M_V \leq -10$) \citep{Kochanek2014}. Events on the bright end such as  NGC 4490 2011OT and M101-OT are expected to be far less common, at most one per century.

\begin{figure*}
\includegraphics[width=0.75\textwidth]{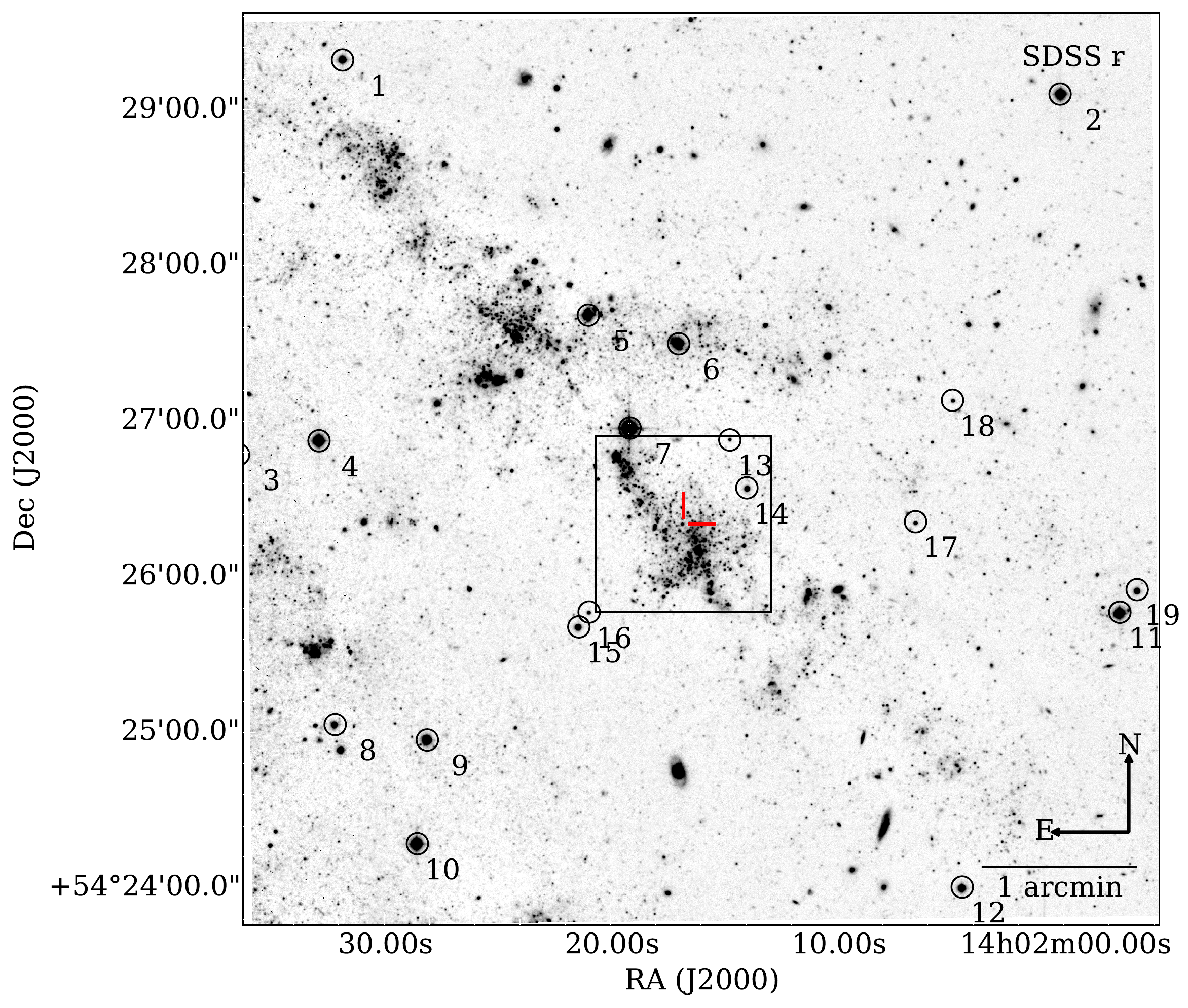} 
\includegraphics[width=0.15\textwidth]{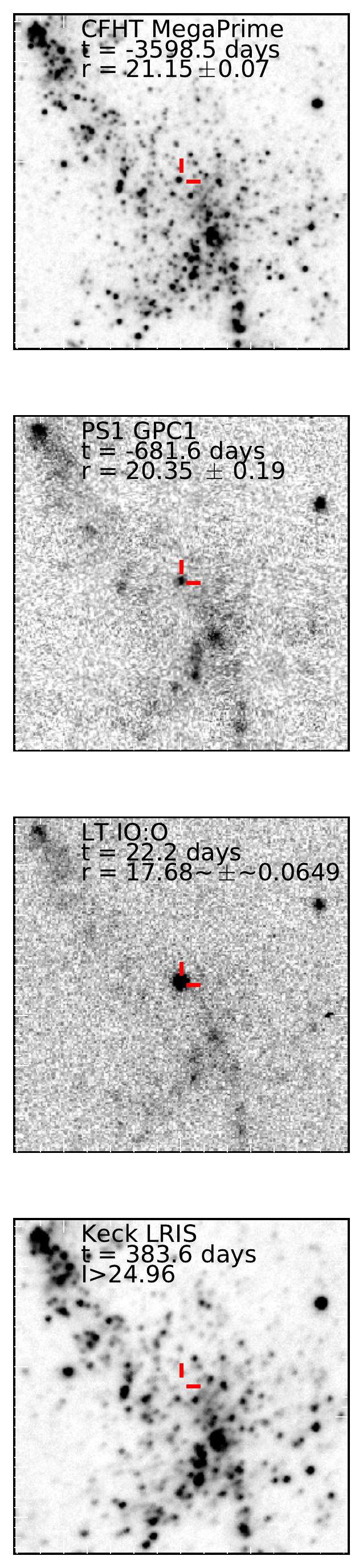}
\caption{Left: M101-OT and the reference stars used to calibrate the photometric zero-point. Due to a variable field-of-view, position and position angle for the M101-OT historical photometric images, different subsets of the fields stars were used according to their visibility. At any time, a minimum number of three stars was used. The square region around M101-OT is shown in detail on the right hand side.
Right: Images of M101-OT at four epochs: $\approx$10\ yrs before reference epoch, 22.3 month, 22 days after the second outburst and 12.6 months after. The field of view size is $1'\times1'$  centred on the position of M101-OT. The red dashes show the location of the transient. The telescope, instrument, and magnitude of the object are listed for each image. }
\label{fig:field}
\end{figure*}


\begin{figure*}
\includegraphics[width=\linewidth]{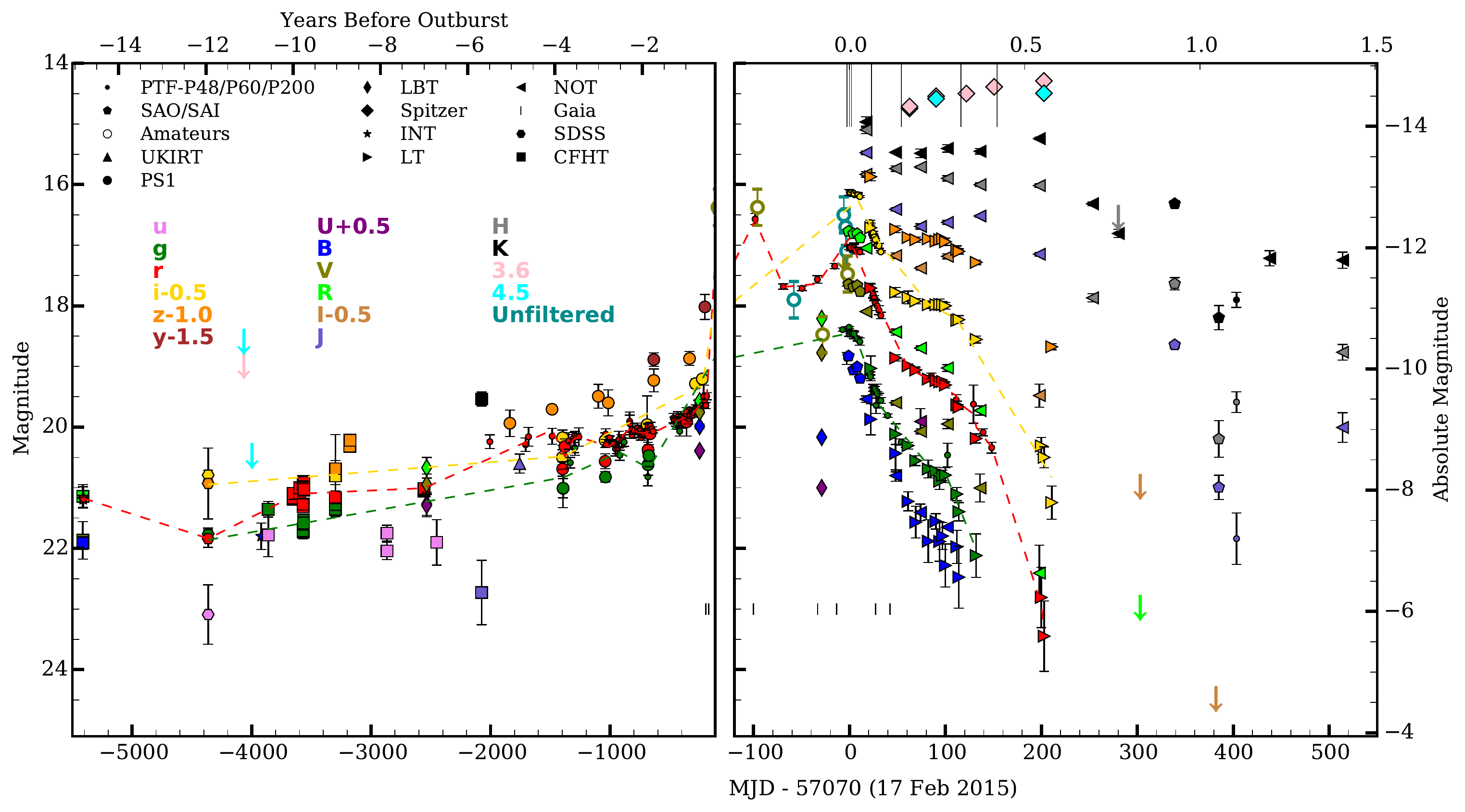} 
\caption{Left: Historic light curve for M101-OT spanning fifteen years of observations until 120 days before second peak date. Pan-STARRS1 and iPTF data allow us to notice an increase in the baseline magnitude of the transient at about 5.5 years before the eruption. Dashed lines are used to guide the eye. Downward-pointing arrows indicate upper limits. Right: Close up of the lightcurve from $-$120 days to +550 days after outburst. For each data point, the marker shape shows the telescope and the colour indicates the filter. Note the difference in time scale between the left and right hand side plots. Vertical tickmarks below the lightcurve show the epochs when this object was observed by \textit{Gaia} (still proprietary data). Upper vertical lines show the epochs when spectra were taken. The lightcurve shows two maxima at $\sim -100$ and 0 days.}
\label{fig:lightcurve}
\end{figure*}



\begin{figure*}
\includegraphics[width=\linewidth]{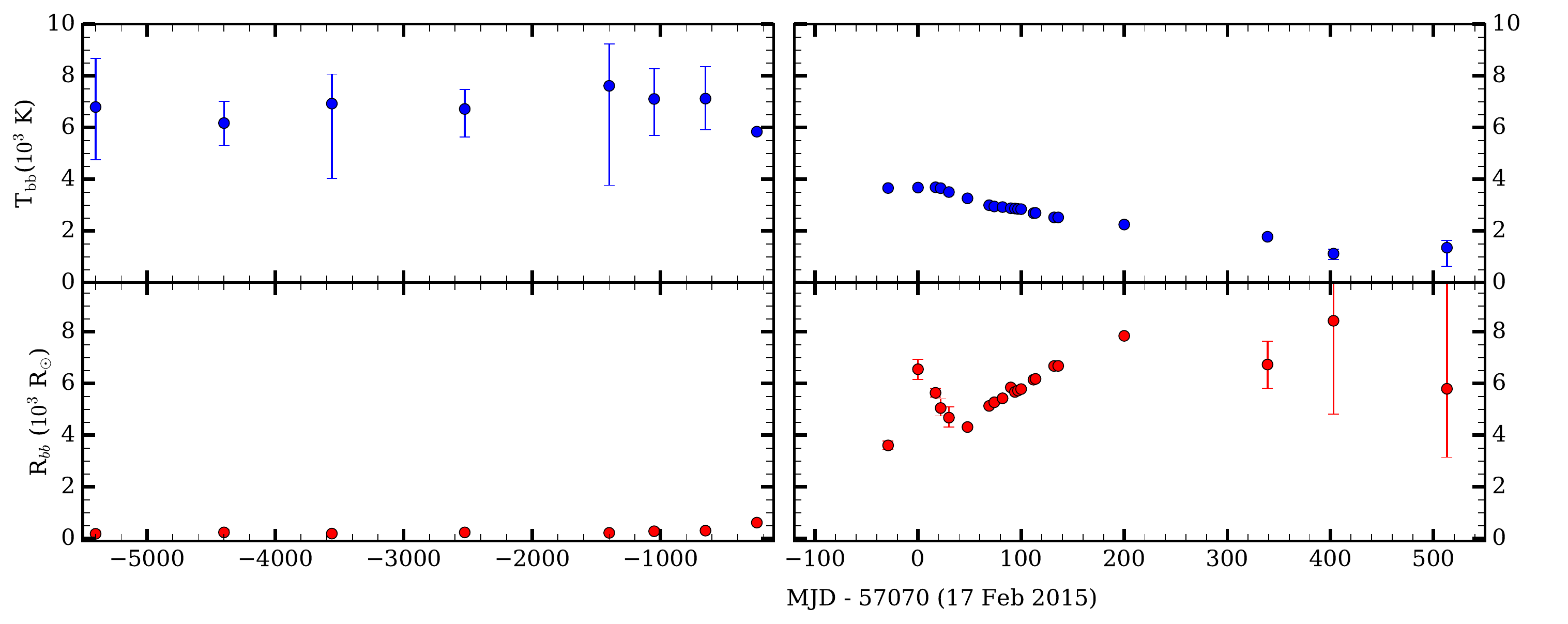} 
\caption{Left: Evolution of the black-body temperature and radius for M101-OT derived from photometry fits for the same time span as the lightcurve. We refer to the analysis in Section \ref{sec:analysis}. Right: Zoom from -120 to +550 days. }
\label{fig:bb_fit}
\end{figure*}

In this work, we will discuss the observations and nature of M101 OT2015-1 (hereafter M101-OT), also designated as PSN J14021678+5426205 and iPTF13afz \citep{ATel7070}, an extragalactic transient in the luminosity gap. The discovery of M101-OT was publicly announced via the IAU Central Bureau for Astronomical Telegrams (CBAT) by Dimitru Ciprian Vintdevara  on the night of 10th to 11th of February 2015 in the outskirts of NGC 5457 (M101) \footnote{ \url{http://www.cbat.eps.harvard.edu/unconf/followups/J14021678+5426205.html} }. Shortly after it was confirmed as an optical transient by Stu Parker with an unfiltered magnitude of 16.7. The source also had an independent discovery within the intermediate Palomar Transient Factory (iPTF) survey back in 2013, when the progenitor was identified as a slow rising source \citep{ATel7070}. This paper is organized as follows: in Section \ref{sec:data}, we report both pre- and post-discovery optical, near-infrared (NIR) and mid-infrared (MIR) photometry and spectroscopy of M101-OT. In Section \ref{sec:analysis}, we examine the spectroscopic measurements and the characteristics of the progenitor. We discuss possible similarities with other objects and the nature of M101-OT in Section \ref{sec:discussion}. Finally, we present a summary and our conclusions in Section \ref{sec:conclusions}.

\section[]{Observations}\label{sec:data}

M101-OT is located ($\alpha_{J2000}=14^{\rm{h}}02^{\rm{m}}16^{\rm{s}}.78\ \delta_{J2000}=+54^{\rm{h}}26^{\rm{m}}20^{\rm{s}}.5$) in the outer reaches of a spiral arm of M101, at
3$'$.41\,N and 8$'$.12\,W of the measured position of the galaxy nucleus. The surrounding region shows signs of a young stellar population, displaying bright unresolved emission in the Galaxy Evolution Explorer (GALEX) survey at 135 nm to 280 nm. 

We adopt the Cepheid distance to M101 of D$_{\rm{L}}~=~6.4~\pm~0.2$\, Mpc, corresponding to a distance modulus of $\mu~=~29.04~\pm~0.05$ (random) $\pm~0.18$ (systematic) mag \citep{ShappeeStanek2011}.
The estimated Galactic reddening at the position of the transient is $E(B-V)~=~0.008~\pm~0.001$\ mag (from NED\footnote{The NASA/IPAC Extragalactic Database (NED) is operated by the Jet Propulsion Laboratory, California Institute of Technology, under contract with the National Aeronautics and Space Administration.} adopting \cite{Schlafly2011}), with R$_{\rm V}~=~3.1$, which corresponds to a mean visual extinction of A$_{\rm V}$ = 0.024 mag. The magnitudes reported in the text and figures of this paper have been corrected for Galactic reddening, but the Tables in the Appendix list the observed magnitudes, i.e. not corrected for extinction.
The extinction within the host galaxy is not included. Local extinction to the progenitor is unlikely, as archival NIR photometry of M101-OT agrees well with the Rayleigh$-$Jeans tail of a single black body emission derived from optical measurements. Therefore, we argue that there is no evidence of a strong warm dust emission component in the environment around the progenitor star.


\subsection{Photometry} \label{sec:phot}

The location of M101-OT has been serendipitously imaged by numerous telescopes and instruments over the last 15 years (from 2000 to 2015). For example, in 2011, this galaxy received special attention, as it hosted one of the youngest SN Ia discovered to date: SN 2011fe \citep{Nugent2011}. In an attempt to piece together the past evolution on M101-OT, we retrieved all available data (see description below) covering the location of the transient. The left panel of Figure \ref{fig:field} shows the location of M101-OT and reference stars used for calibration. The right panel shows the magnitude evolution for $-$10 yr, $-$1.8 yr and an early follow-up epoch at 22 days after the second peak. The source has faded below detectable limits at +383 days. Throughout this work, we will use as a reference epoch the date of the second peak in $r$-band, MJD~57070.

Our best quality pre-discovery image (seeing of 0$\farcs$55) is an $r$-band exposure at $-$3625 days pre-peak from the Canada France Hawaii Telescope (CFHT). We aligned this image with our +22-d post-peak image using 18 stars in common. There is one point source (see right hand side of Figure \ref{fig:field}) in the image within a 2$''$ radius of the position during the outburst, and the central position of the point spread functions (PSFs) are coincident within 180 mas (with a precision in the alignment of 250 mas). We identify this point source as the progenitor of M101-OT. Imaging in $I$-band taken at late times with Keck confirm the disappearance of the progenitor star.

The historical optical data for M101-OT was retrieved from the CFHT MegaPrime and CFHT12K/Mosaic, using single and combined exposures \citep{Gwyn2008}, Pan-STARRS-1/GPC1 \citep[][PS1;]{Magnier2013,Schlafly2012,Tonry2012}, Isaac Newton Telescope/Wide Field Camera (INT/WFC) and Sloan Digital Sky Survey (SDSS) DR 10 \citep{SDSS10}. Unfortunately, there are no Hubble Space Telescope (HST) images covering the location of the source. Post discovery optical magnitudes were obtained from the reported follow-up astronomer\'s telegrams (ATels), Liverpool Telescope (LT), the Nordic Optical Telescope (NOT) and the Palomar P48 and P60 telescopes. The infrared data were retrieved from CFHT/WIRCam, UKIRT/WFCAM and the Spitzer Infrared Array Camera \citep{Fazio2004} in 3.6 and 4.5 $\mu$m as part of the SPitzer InfraRed Intensive Transients Survey (SPIRITS) (Kasliwal. et. al. in prep). Details of pre-discovery photometry and post-discovery optical photometry may be found in the Appendices Table \ref{table:preoutburst} and Table \ref{table:followup} respectively. iPTF photometry is reported in Table \ref{table:ptf}. The NIR and MIR observations are summarized in Appendix Table \ref{table:irphot}. 

\begin{figure}
\centering
\includegraphics[width=0.5\textwidth]{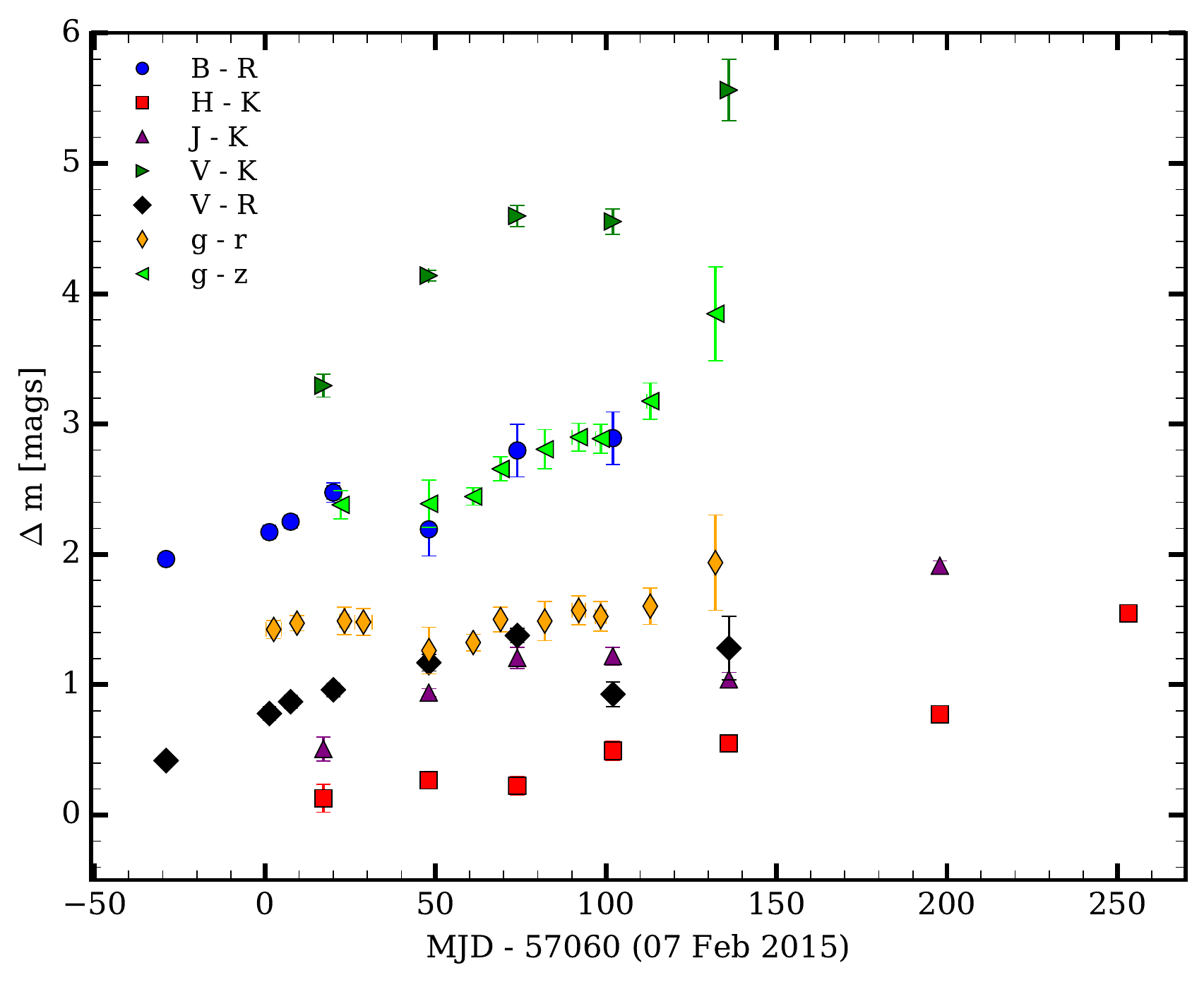} 
\caption{Post-outburst colour evolution for M101-OT. The data points have been binned in groups of 10 days. The abscissa is the average MJD of the bin relative to the reference epoch. }
\label{fig:colour_evolution}
\end{figure}

\begin{figure*}[h]
\centering
\includegraphics[width=0.82\linewidth]{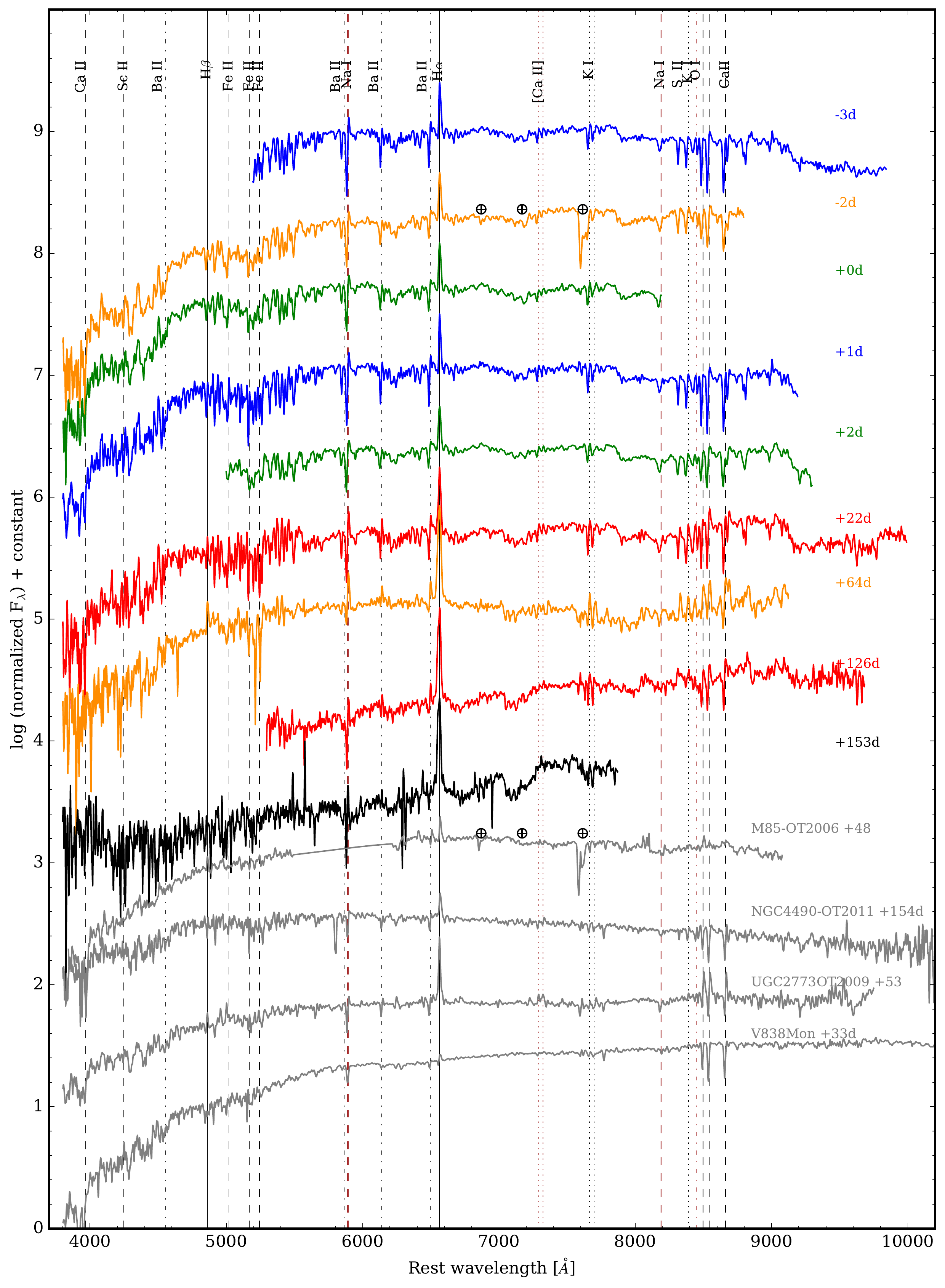} 
\caption{Spectral evolution of M101-OT. The spectra have been flux calibrated using the interpolated photometric measurements. Telluric absorption features were corrected or marked otherwise with $\oplus$ symbol. Because the response of the detector drops at the extremes, some spectra are only shown for the valid wavelength range. The spectrum is colour coded by instrument. Blue: P200+DPSP, green: 1.82m+AFOSC, red: WHT+ISIS, orange: NOT+ALFOSC, black: GTC+OSIRIS, grey: comparison spectra. The Keck/LRIS spectrum of M85-OT is from \citep{Kulkarni2007}, NGC 4490-OT from \citep{Smith2016}, the UGC2773-OT2009 spectrum was taken with TNG/DOLORES (A. Pastorello, private communication), and finally the spectrum of V838 Mon was taken with the Kast spectrograph at Lick Observatory on 2002 Mar. 11 \citep{Smith2016}.}
\label{fig:spec}
\end{figure*}

We measured the brightness of the source coincident with M101-OT using the \textsc{IRAF} (Image Reduction and Analysis Facility) SNOoPY \footnote{SNOoPY is a package developed by
E. Cappellaro, based on \textsc{daophot}, but optimized for SN magnitude measurements.} package for PSF photometry. The zero-point in the SDSS photometric system was calibrated using aperture photometry on three to nineteen different sequence stars in the M101-OT field. Figure \ref{fig:field} shows the position of the sequence stars. Their coordinates and magnitudes are reported in Table \ref{table:seq}. The magnitude measurements for bands \textit{grizy} were obtained from the PS1 catalogue, having photometric accuracy better than 0.01 mag. Measurements for \textit{u}-band were obtained from the SDSS DR10 catalogue. Johnson photometry was calibrated using the same PS1 catalogue and transformations provided by \cite{Tonry2012} with a root-mean-square (RMS) below 0.1 mag. Photometry from the iPTF survey was obtained with the Palomar Transient Factory Image Differencing Extraction
(PTFIDE) pipeline for the 48-inch telescope \citep{Masci2016} and with a custom difference imaging pipeline for 60-inch telescope at Mount Palomar \citep{Cenko2006}. Transformations from PTF Mould-$R$ and $g$-band to SDSS equivalent photometry were obtained using the transformations in \cite{Ofek2012}. The NOT NIR reductions were based on using an external \textsc{IRAF} package \textsc{notcam version 2.5}\footnote{ \url{http://www.not.iac.es/instruments/notcam/guide/observe.html}} and a custom pipeline for WIRC data. The zero-point for IR photometry was calibrated using the  Two Micron All Sky Survey (2MASS) photometry. 

The full historical lightcurve of M101-OT is shown in Figure \ref{fig:lightcurve}, left panel. The earliest detection of the progenitor was obtained on 2000 February 05 with CFHT. From these first single epoch observations, we get $B=21.9\pm$0.1, $V=21.8\pm$0.3, $R=21.1\pm$0.2 (corrected for Milky Way reddening), which at the distance of M101 yield an absolute magnitudes of $M_B=-7.1$, $M_V=-7.2$, and $M_R=-7.9$ for the progenitor star. INT observations in the $r$-band, $r=21.2\pm0.15$, taken only three days after CFHT, are consistent within 0.1 mag with the $R$-band measurements. 

Within the first period, from approximately 15 to 5.5 years before outburst, the brightness of the progenitor shows only minor variations. The magnitude in the $r$-band remained constant to within 0.2 mags, with an average value of $r$=21.1. Roughly 5.5 years before the outburst, the lightcurve began to rise smoothly across all bands. The $r$-band increased to 19.6 mag at $-$180 days i.e. 1.5 mag relative to the historical median value. Reported magnitudes prior to mid-2012 ($-$2.7 years) taken with the Large Binocular Telescope (LBT) agree with these values: the source was reported as being variable, with mean magnitudes of $U=21.33\pm$0.19, $B=21.30\pm$0.19, $V=20.97\pm$0.17, and $R=20.69\pm$0.17 \citep{ATel7069}. During its slow rise, the transient was detected on 4th of April 2013 (-684 days) internally by iPTF as a slowly brightening source.

On 2014 November 10, after appearing from behind the Sun, it was detected at 16.6 mag in R-band during its first outburst \citep{ATel7070}, $\sim$3 months prior to public discovery. At approximately -29 days to peak, it was also detected by LBT in between the first and second outbursts at a considerably fainter magnitude of $R~=~18.22\pm$0.02 \citep{ATel7069}. At the time of public announcement on 2015 February 10, the object was close to its second peak, estimated to fall 10 days later, on 2015 February 17 (MJD 57070). 

The Gaia satellite \citep{Perryman2001} (a European Space Agency mission) serendipitously observed the region containing M101-OT
during the time of the first peak. Unfortunately, these data have not been made available to us.
Due to this handicap in constraining the time of the first peak, we choose to adopt the epoch of maximum brightness of the second peak, at MJD 57070, as our reference epoch.

The follow-up photometry for M101-OT is shown in the right panel of Figure \ref{fig:lightcurve}. The most remarkable feature of the lightcurve is the existence of two maxima. The object was observed during the decay phase of the first peak, having an absolute magnitude of $M_r\ \leq -12.6$ mag (we only have data on the decline part for the first peak, so the outburst could have been brighter). The second maximum, $\sim 100$ days after shows $M_r\ \simeq -12.0$ mag and it is followed by a fast declining phase, lasting $\sim$40 days, when the object fades 2 magnitudes in $r$-band. The lightcurve makes a transition into a plateau phase of $\sim$60 days: the redder $riz$-bands flatten, while the bluer $Bg$-bands continue to decline. After the end of the plateau, around +110 days, the transient resumes the initial decline rate in $r$-band. The first NIR follow-up data show magnitudes of J=15.45 $\pm$ 0.3, H=15.07 $\pm$ 0.06 and K=14.94 $\pm$ 0.09 at +17 days. The evolution in the IR bands is slow, and only after day +200 the object starts to decline in the IR too. Between +200 and +256 days it fades by $\sim$1 mag in the $K$-band. However, later epoch observations provided by \citep{ATel7206} and follow-up with P200 and NOT, suggest a re-brightening of the object in IR bands. Multi-band photometry allows us to derive the black body temperature and radius of the object, shown in  Figure \ref{fig:bb_fit} (see \ref{sec:blackbody} for details).

The colour evolution between $-$29 and +272\,d for M101-OT is shown in Figure \ref{fig:colour_evolution}. Coincident with the end of the first phase of the lightcurve, at $\sim 50$ days, the object becomes slightly bluer in $B$ and $g$-bands. This period is associated with a decrease of the photospheric radius. At approximately +130 days, around the end of the plateau, the colour evolution shows a second temporary ($\sim$20 days) enhancement of flux ratio for the blue bands. The last multi-band epoch (+154 days) shows that the object becomes increasingly red, i.e. $g-r=1.9\ \pm\ 0.4$, $g-z=3.8\ \pm\ 0.4$ and $V-K=5.6\ \pm\ 0.2$.

 \subsection{Optical Spectroscopy} \label{sec:spec}

\begin{figure}
\centering
\includegraphics[width=0.5\textwidth]{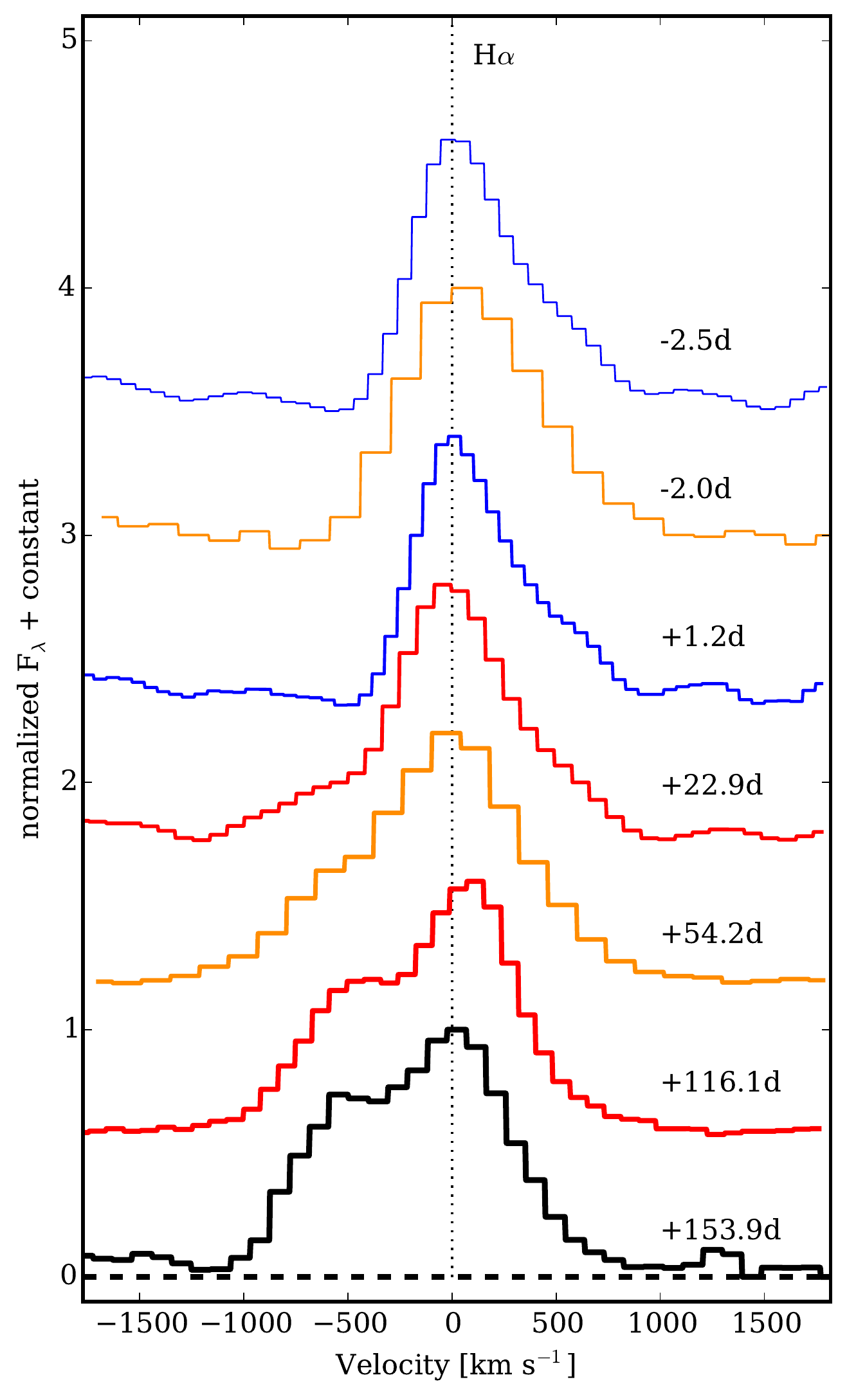} 
\caption{The continuum-subtracted and peak normalized H$\alpha$ region for DBSP (500 \kms), WHT (320 \kms), NOT (747 \kms) and GTC (380 \kms) spectra. The spectrum is colour coded by instrument, same as in Figure \ref{fig:spec}. }
\label{fig:halpha}
\end{figure}

We obtained spectra of M101-OT using a range of facilities. The log of the spectroscopic observations
is given in Table \ref{table:speclog}. The data were reduced using \textsc{IRAF} and \textsc{PyRAF} standard routines. The wavelength calibration was done by fitting low-order polynomials to arc lamp spectra. Sky lines were used to check the accuracy of the calibration, which is within 1 \angstrom. We calibrated the flux using spectro-photometric standard stars and later on adjusted it using interpolated photometry for the same epoch as the spectra. The spectra of M101-OT are made public via WISeREP \citep{YaronGal-Yam2012}.

We assumed the heliocentric recessional velocity for M101 of 241 $\pm$ 2 \kms \citep{deVaucouleurs1991}. Figure \ref{fig:spec} shows the spectral evolution of M101-OT.  All spectra show a cool photospheric continuum, fitted by a black-body emission with temperatures 3000 $-$ 3600 K. 

 The blue part of the M101-OT spectrum is dominated by the absorption forest of Fe~II (at around 5400 \AA), Ti~II (below 4700 \AA)  and Sc~II lines. P-Cygni profiles are displayed by intermediate-mass elements. Ca II is identified with an expansion velocity of $v\ \simeq-356\ \pm\ 9$ \kms for the absorption component at +2 days, slowing to $v\ \simeq\ -283\ \pm\ 2$ \kms at +22 days, and $v\ \simeq-207\ \pm\ 17$ \kms at +116 days. Ba II $\lambda \lambda$ 6134, 6489 is also identified as P-Cygni profile with an expansion velocity of $-$180 \kms and a full width at half maximum (FWHM) of 367 \kms. Other elements present in early-time spectra are $\lambda \lambda$ 5150 Mg $\lambda \lambda$ and Na I at $\lambda\lambda$ 5890, 5896 and $\lambda\lambda$ 8183, 8195.  Resonance lines K I $\lambda \lambda$ 7665, 7699 and $\lambda \lambda$ 7665, 7699 are also found in the spectrum, although their P-Cygni profiles are much weaker. These lines are rare and have been seen in the extreme super-giant VY CMa \citep{Smith2004MNRASVY} and  Type IIn SN 2009kn \citep{Kankare2012}. We do not detect strong [Ca II] $\lambda\lambda$7291, 7325 lines in the spectrum, which have been associated with dense and compact gas disk and presence of dust \citep{Smith2010,Liermann2014}. Figure \ref{fig:spec} shows comparison spectra for similar red transients. M85-OT2006-1 is defined as an SN2008S-like observational class, showing strong emission for CaII and [Ca II] lines. UGC2773-OT2009-1 is considered to be an example of a dust enshrouded LBV. NGC4490-OT2011-1 and V838 Mon are examples of LRNe. There is an important resemblance between all three groups, implying that the nature of the outburst can not be determined from spectra alone.

The spectra of M101-OT have a significant evolution of the H$\alpha$ profile. Figure \ref{fig:halpha} shows different morphologies of the profiles for different epochs. At early times, its expansion velocity, derived from the FWHM, is around 500\kms, slightly larger than the one of intermediate mass elements. The profile is asymmetric and shows a small blueshifted absorption component. However, at +22.9 days the absorption evolves into an emission profile, suggesting the existence of asymmetry in the outflow. The implications of this are further discussed in Section \ref{sec:halpha}. Similar behaviour was observed in the high resolution spectra of NGC4490-OT2011-1 reported in \cite{Smith2016}.

\section{Analysis}\label{sec:analysis}

\subsection{Spectroscopic Analysis}
\subsubsection{The H$\alpha$ profile}
\label{sec:halpha}
An interesting feature is the evolution to a double peaked profile of the H$\alpha$ line (Fig. \ref{fig:halpha}). There is evidence for a double peaked line, with a difference in velocity of $\sim$500 \kms between the components. Spectra taken around peak show the blueshifted P-Cygni component in absorption.

However, for later epochs, after the beginning of the plateau phase at +40 days, the absorption disappears and an increasingly bright blueshifted emission peak appears instead. The second emission component becomes clearly visible at +54 days, and reaches similar equivalent width as the redshifted counterpart at +116 days. 

Similar absorption in the blue wing evolving into an emission component was also observed for the LRNe V1309 Sco \citep{Mason2010} and NGC 4490 2011OT-1 \citep{Smith2016}, which both had higher spectral resolution data.

\subsubsection{Molecular bands}
\label{sec:molecular_bands}

\begin{figure}
\centering
\includegraphics[width=0.5\textwidth]{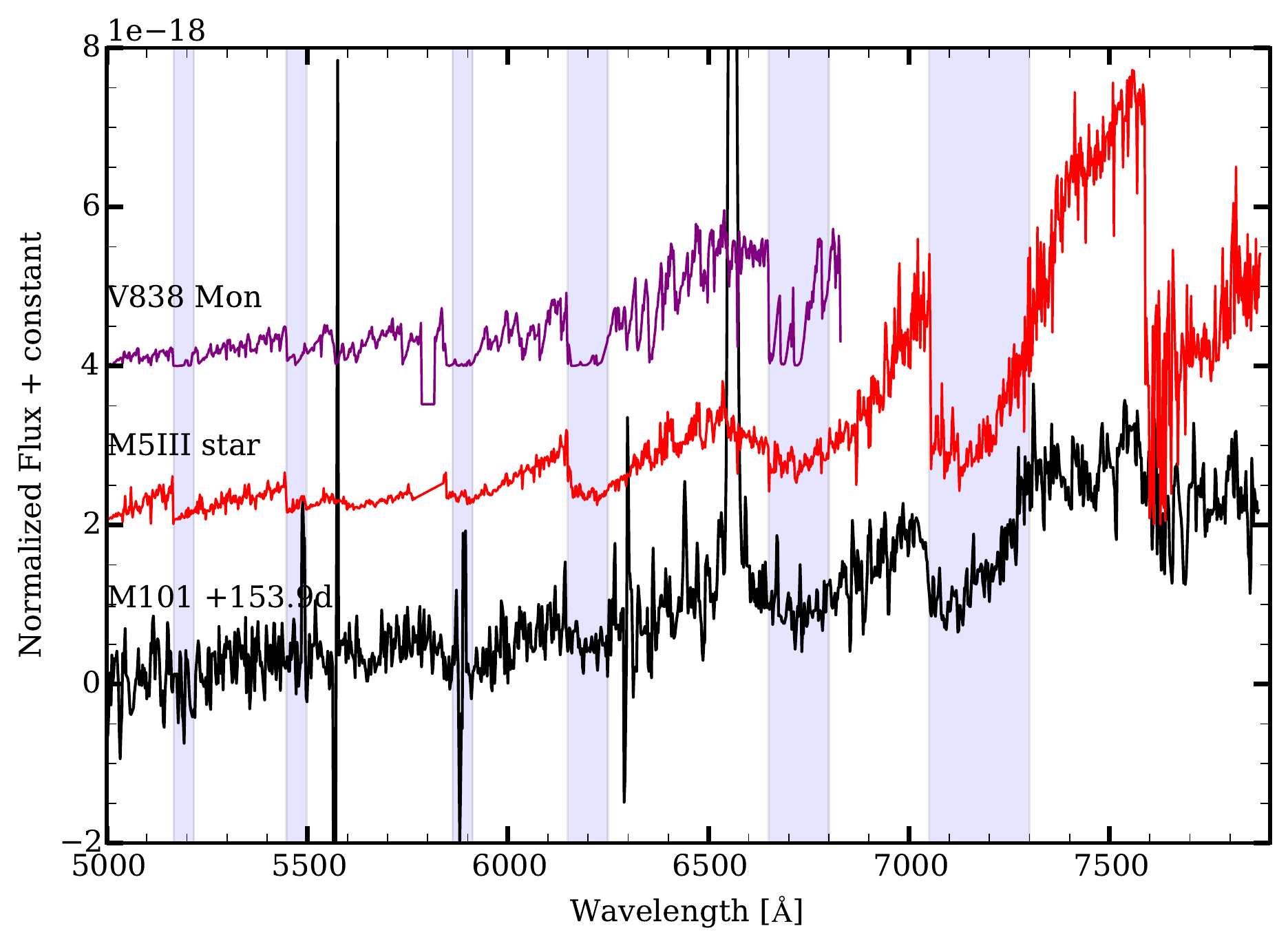} 
\caption{ Comparison of M101-OT spectrum at +154 days with HD118767 M5III star\citep{Bagnulo2003} and the average spectrum of V838 Mon \citep{Tylenda2011}. The spectrum has an estimated black-body temperature of $\sim$3000 K. The molecular bands are comparable to the ones in cool giant stars. Major molecular absorption lines are marked in the spectrum with blue vertical bands.}
\label{fig:m5iii}
\end{figure}

Spectra taken at +116 days and later epochs show the initial formation of molecular bands, characteristic for cool M-type stars. Figure \ref{fig:m5iii} shows the comparison between M101-OT spectrum at +154 days, with a cool M5III star and the Galactic merger V838 Mon, seven years after its outburst. At this phase, the photospheric temperature shows a good fit with $\sim 3000$ K black-body. We detect titanium oxide (TiO) bands in the range 6600-6800 \AA\ and 7050-7300 \AA. Between 7300 and 7600 \AA, TiO absorption is combined with vanadium oxide (VO) molecular absorption, which becomes dominant above 7400 \AA.

\subsection{SED analysis and bolometric lightcurve} \label{sec:blackbody}

We computed a black-body fit to several pre- and post- discovery epochs, preferentially taken around the same epoch, or at most $\pm$50 days from each other. In the case that a particular band had more than one measurement within the time interval, we computed the mean value weighted by the errors. 

We used the MCMC \textsc{python} package \textsc{emcee} \citep{Foreman-Mackey2013} to obtain the value of the maximum posterior probability and 1$\sigma$ confidence intervals on the estimated parameters. The evolution of temperature and radius for the best black-body fit is shown in Figure \ref{fig:bb_fit}.
In all cases, a single black-body component was sufficient to describe the observed spectral energy distribution.

The initial fits for the progenitor at epochs earlier than 6 years, show that the temperature and radius were constant within the errors with values of $T=6600\ \pm$ 300 K and an $R=220\pm25$ \Rsun. Starting at $-$5.5 years, there was a progressive expansion and cooling of the star, so that at $-$250 days it cooled down to $T=5800\pm$120\,K and nearly tripled its radius to $R=620\pm25$ \Rsun. During the peak of the second outburst, the temperature had decreased to roughly 3300 K, and continued to cool down slowly over the next 400 days. The photospheric radius showed a peculiar behaviour. It had grown exponentially up to $R\ \sim 6500 \pm 400$ \Rsun during outburst peak, receded to $R\ \sim 4300 \pm 80$ \Rsun at 48 days and expanded again to approximately $R\ \sim 7800 \pm 50$ \Rsun at 200 days. A similar effect was noticed for M31 2015 LRN \citep{MacLeod2016}. We fit a linear model for the radial expansion for epochs 70 to 200 days, which allowed to derive the photospheric expansion velocity of 170$~\pm~$5 \kms. 

Around the second outburst, from $-$30 days to 20 days, the temperature had a constant value of 3670 $\pm$ 50 K. The plateau phase, detected in the redder bands, is associated with a slower decline in the temperature: $\simeq$150 K between days 40 and 100.  IR photometry for later epochs ($>$ 400) show that the temperature is consistent with 1200 $\pm$300 K black-body emission.

\begin{figure}
\centering
\includegraphics[width=0.5\textwidth]{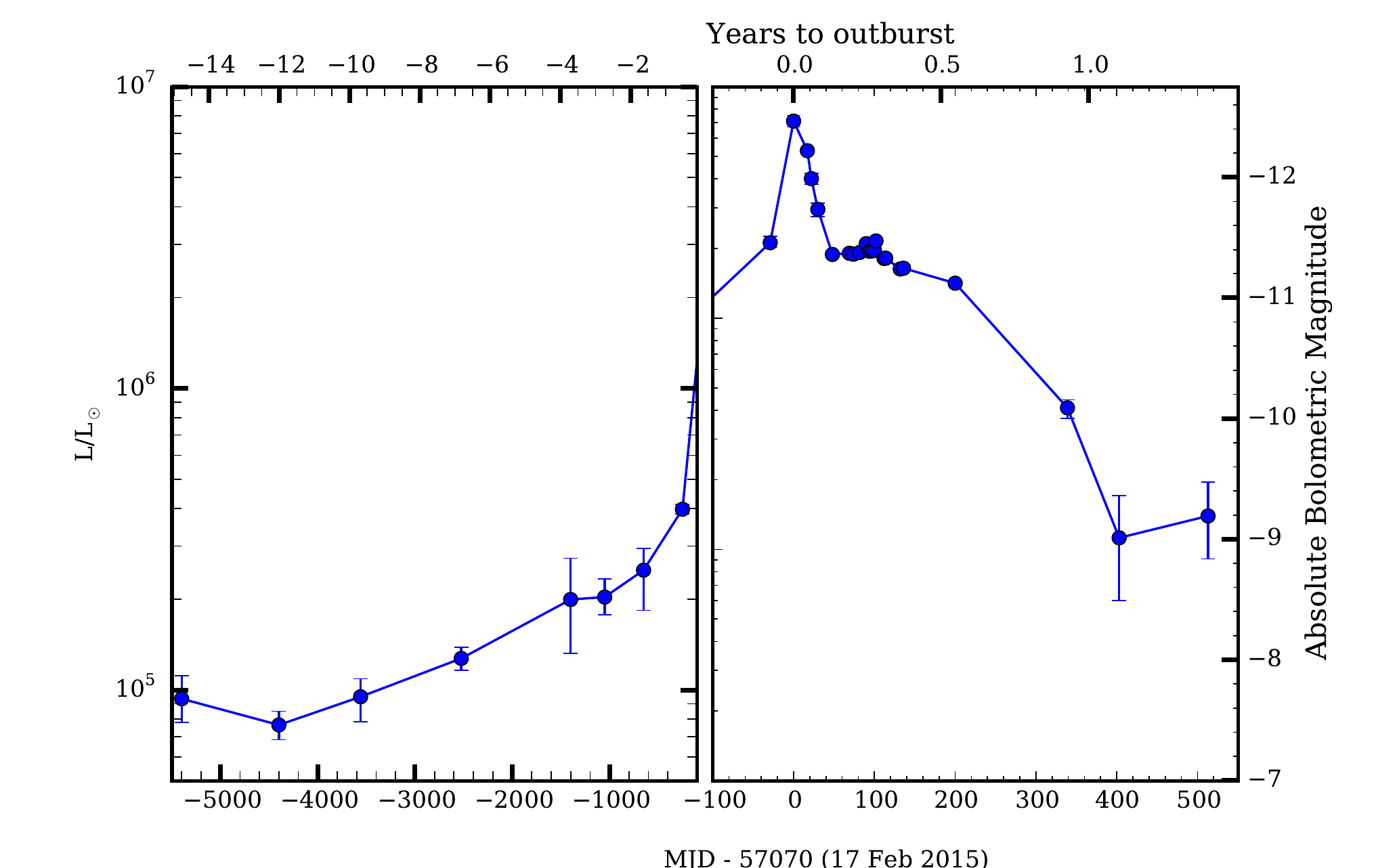} 
\caption{The evolution of the black-body luminosity of M101-OT. The first peak is not present, as there were not enough photometric measurements for a reliable fit. The measurements for epochs after +200 days were derived using NIR bands only. }
\label{fig:lbol}
\end{figure}

The integrated black-body emission was used to estimate the bolometric lightcurve for M101-OT, shown in Figure \ref{fig:lbol}. While the early time photometry shows a rather stable object with luminosity $L\ \simeq 2.6\ \times 10^5$ \Lsun, photometry later than five years prior to the outburst shows a steady increase in the star's bolometric luminosity, reaching $L\ \sim\ 4 \times 10^5$ \Lsun at 250 days before and approximately $L\ \simeq 6.3\ \times 10^6$ \Lsun during the maximum of the second outburst. 

\subsection{Progenitor analysis}

\subsubsection{Single star scenario}

Photometric measurements from the earliest three archival epochs, obtained between 15 and 8 years before outburst, were used to derive the best parameters for the progenitor star. We found a good agreement with a single black-body fit. No significant IR excess was observed in the early photometric measurements. The star was estimated to have a temperature of $T=6600\ \pm$ 300 K and an approximate radius of  $R=220\pm13$ \Rsun. The historic average bolometric luminosity is $L \sim 8.8~\pm~0.8\times 10^4 $ \Lsun, which placed the progenitor star to be below the low luminosity end of the LBV zone in the HR diagram \citep{SmithVink2004ApJ}, where known LBVs tend to have luminosities greater than $L \sim 2\ \times 10^5 $ \Lsun.

In order to derive the characteristics of the progenitor system, we compared (using maximum likelihood) the observed broad band photometric archival measurements with the predicted absolute magnitudes in the BPASS models \footnote{\url{http://bpass.auckland.ac.nz}}.  Specifically, we obtained the averaged photometric measurements over all epochs older than -5.5 years to compare with the predicted photometry of the system for both single and binary stellar models. Single star evolution tracks were taken from BPASS v1.0  \citep{Eldridge2004} and binary stellar evolution tracks from BPASS v2.0 \citep{Stanway2016}. We assumed Solar metallicity for in both cases.

\begin{figure*}
\centering
\includegraphics[width=0.51\textwidth]{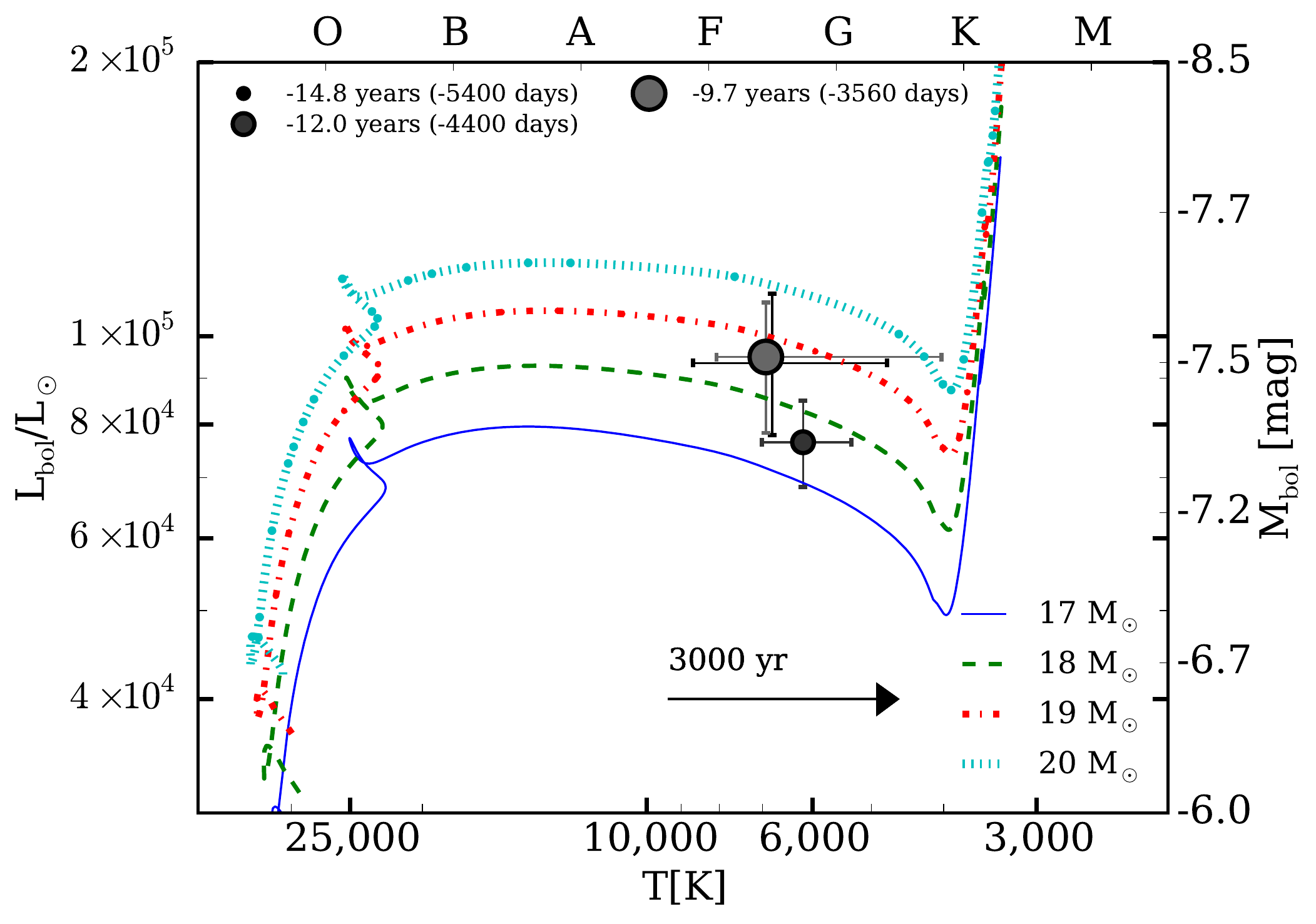} 
\hspace{-0.3cm}
\caption{The position of the progenitor of M101-OT for the first three initial epochs older than 2500 days, when L$_{\rm bol}$ is constant. The size of the point encodes the epoch of the observation relative to discovery. Stellar evolutionary tracks for single star models \citep{Eldridge2004} are shown for stars with initial masses from 17 to 20 \Msun. To provide a graphic example of the short life of stars in the progenitor region, the arrow shows the temperature locus covered by an 18 \Msun star in only 3000 years.}
\label{fig:stellar_tracks}
\end{figure*}

For the case of a single star evolution scenario with fixed metallicity, the only free parameter of the model is the initial progenitor mass. Figure \ref{fig:stellar_tracks} shows the location of the progenitor star in the temperature-luminosity space for three historic measurements. The progenitor star is consistent with an F-type yellow super-giant, with initial mass of 18-19 \Msun, that is evolving off the main sequence towards the red super-giant (RSG) phase. The location of the progenitor, named the Hertzsprung gap, is extremely unusual, as it is associated with stars that finished core hydrogen burning, but have not started yet the shell hydrogen burning phase. Stars are expected to spend only a small fraction of their lives ($\sim$ 3000 yr) in the region where the progenitor system is found, as shown in Figure \ref{fig:stellar_tracks}. Stars in the gap experience an exponential increase in the stellar radius, from $\sim$20 \Rsun to $\sim$800 \Rsun for a 18 \Msun star. The age of the star is when the progenitor reaches 230 \Rsun is 9.9$\pm$0.1 Myr.

\subsubsection{Binary star scenario}

Detailed modeling of the event assuming a binary star evolution scenario is beyond the scope of this paper. In the current work, we aim to provide initial constraints on the progenitor system and the possible fate of the remnant.

We define the common envelope (CE) evolution as a short lived phase in the evolution of an interacting binary system \citep{Paczynski1976}. It is initiated when the most evolved star expands enough to overfill its Roche lobe (RL), triggering an unstable mass transfer towards its companion, which accumulates in a common envelope surrounding both stars (see \cite{Ivanova2013Rev} for a review). The Roche lobe radius for the primary star is well approximated by \cite{Eggleton1983}. 

\begin{equation}
RL_1 = a \frac{0.49 q^{2/3}}{0.6 q^{2/3} + \rm{ln}(1 + q^{1/3})}
\end{equation}

Where $q=M1/M2$ is the mass ratio where $M1$ is the mass of the primary, more massive star, and $M2$ is the mass of the secondary; $a$ is the separation between the two components. We assume that the stars are in circular orbit and that the expansion of the primary takes place on a longer timescale than the formation of a CE. Therefore, during the stable phase before the outburst, the radius of the primary will be equivalent to its Roche lobe radius ($R1=RL1$). From our observables, we estimate the radius for the primary star to be R$\simeq$ 230 \Rsun. Under the condition of Roche lobe overflow (RLOF), we can constrain the initial orbital separation $a_i$ for different mass ratios of the system. For example, for a system with nearly equal masses and $q \simeq 1$, this value approximates to $\approx$ 600 \Rsun and for $q=18$ ($M1=18$ \Msun and $M2=1$ \Msun), to $\approx$ 370 \Rsun. The periods associated to these separations are $\approx$290 and $\approx$190 days respectively.

The outcome of the CE phase can be expressed following the basic energy formalism  \citep{deKool1990,Ivanova2013Rev}:

\begin{equation}
G \frac{ M1 M1_{\rm{env}}}{\lambda \rm{R1}} = \alpha_{CE} \left(
-G\frac{M1 M2}{2 a_i} + G \frac{M1_{\rm{c}} M2}{2 a_f} \right)
\end{equation}

\begin{figure*}
\centering
\includegraphics[width=0.51\textwidth]{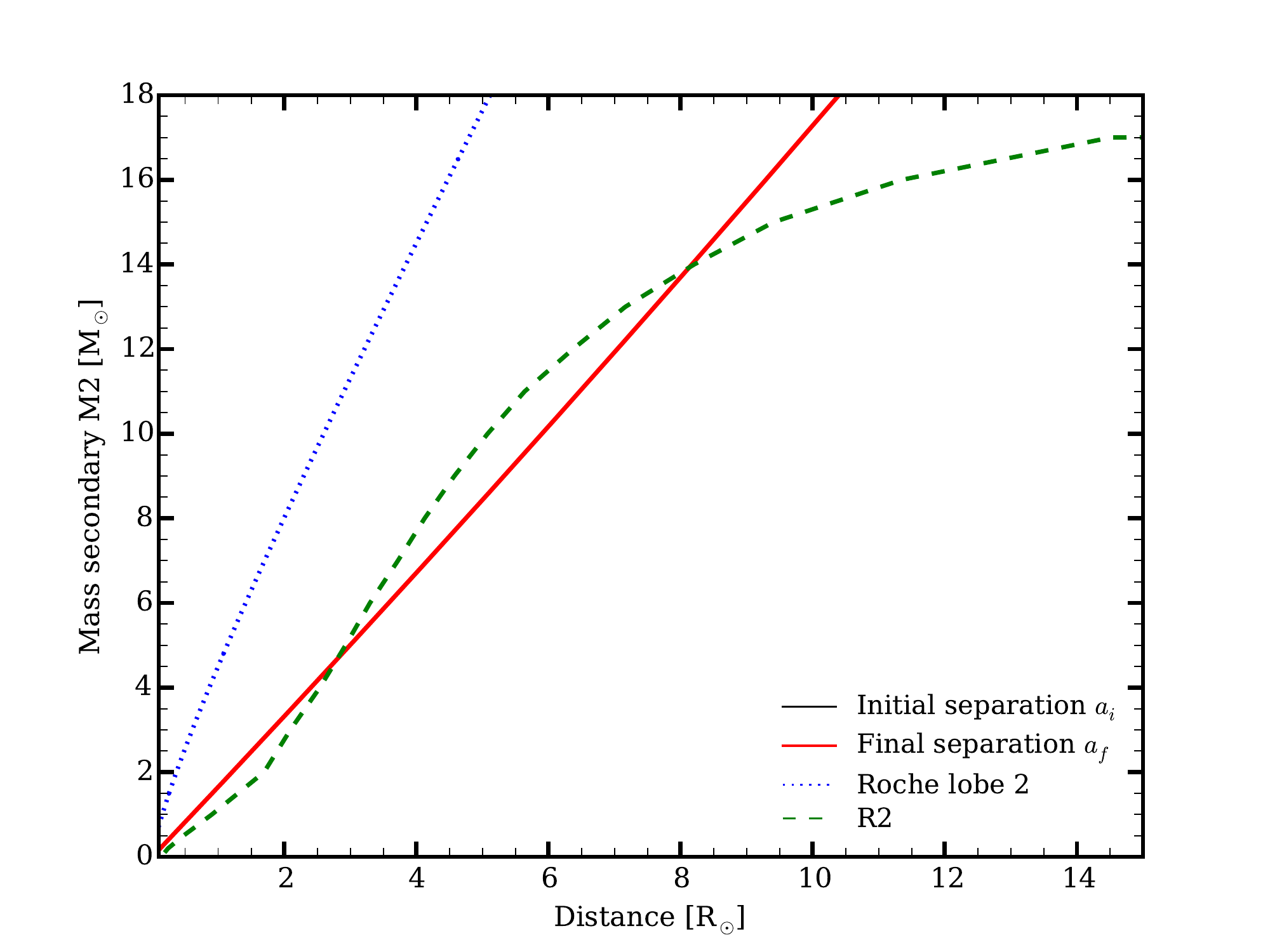} 
\hspace{-0.3cm}
\caption{Final state of the binary system assuming that all the envelope is ejected at the expenses of the orbital energy, reducing the separation of the components from $a_i$ to $a_f$. The primary star has an initial mass of 18\Msun and an envelope of 13\Msun. The initial separation has been estimated from the radius of the primary (see text). The red solid line indicates the boundary final separation, $a_f$, when the release of orbital energy equals the binding energy of the envelope for a given secondary mass $M2$. The blue dotted line shows the Roche lobe of the secondary star for the given $a_f$, secondary mass $M2$ and the core of the remnant $M1_c$=5 \Msun. The dashed green line shows the radius of the secondary star at 9.8 Myr.}
\label{fig:energies}
\end{figure*}

The left side of the equation describes the binding energy of the envelope in the primary star. The right side describes the orbital energy released by the system from its initial orbital separation $a_i$, to its final separation $a_f$ after the loss of the envelope. $G$ is the gravitational constant, $M1_{\rm{env}}$ is the mass of the expelled envelope and $M1_{\rm{c}}$ is the mass of the remaining core. The parameter $\lambda$ is related to the internal envelope structure of the star, and $\alpha_{CE}$ represents the fraction of the gravitational binding energy that is used to eject the envelope with velocities larger than the local escape velocity. We assume $\lambda=0.5$ \citep{deKool1990} and $\alpha_{CE}$ to be 0.5, which accounts for the need of kinetic energy of similar order of magnitude as the binding energy. According to the results derived from single stellar models, we fix $M1$=18 \Msun, $M1_{\rm{env}}=13$ \Msun and $M1_{\rm{c}} = 5$ \Msun. The estimated binding energy for the envelope when the radius of the primary is $R1=230$ \Rsun is of the order of E$_{\rm{bind}}$$\sim 8 \times 10^{48} $erg. In order to be able to eject this envelope completely, the release of orbital energy needs to be equal or larger than E$_{\rm{bind}}$. Figure \ref{fig:energies} shows the parameter space for the mass of the secondary component and the final orbital separation which satisfies the energy balance stated above. We find that, in order to eject the envelope completely, the final separation of the system would need to be of the order of the radius of the secondary star (assuming it has the same age as the primary). The \textit{spiral-in} phase would continue until the final separation $a_f$ have shrunk below the radius of the secondary, eventually leading to the merger of the system. The conclusions presented here are not sensitive to small variations in the mass of the primary ($\pm 1$ \Msun), the initial separation $a_i$, or the dimensionless parameters $\lambda$ and $\alpha_{CE}$, sometimes treated as \textit{fudge} factor in binary population synthesis models. According to these simple calculations, a (nearly) full ejection of the envelope for M101-OT would lead the system to merge. 

An alternative interpretation is provided by binary evolution codes. We have examined the evolution of possible progenitor systems predicted by the BPASS v2.0 binary evolution models. These models, assuming Solar metallicity, have three main parameters: mass of the primary, system mass ratio, and the logarithm of the period. These parameters are sampled in steps of 1 \Msun, 0.2, and 0.2 dex respectively. As before, we assumed an initial mass for the primary $M1=18$\Msun and imposed the constraint on mass ratio and periods derived earlier from the progenitor radius ($R1=RL1$). In all cases, the models predicted a surviving binary system with large final separation for the system, in the range $260 \leq a_f \leq 290$ \Rsun. The discrepancy with the basic energy formalism is likely to be caused by the simplicity of our initial approach, which omits additional sources of energy, such as the star internal energy, the thermal energy of the gas, or the recombination energy. The inclusion of these terms may reduce the magnitude of the binding energy, allowing the binary to survive. We note that the interpretation of these results is only a suggestion and further analysis is needed to draw firmer conclusions.

\section{Discussion} \label{sec:discussion}

The absolute magnitude for M101-OT with peaks at $M_r\leq -12.4$ and $M_r\simeq-12.0$mag, and its red colour, $g-r=1.4$ mag during the secondary peak, places this event in the so-called ``gap'' region of the timescale-luminosity diagram between novae ( $-4$ to $-10$ mag), and SNe ($-15$ to $-22$ mag). Photometrically, the double-peaked lightcurve of M101-OT and increasingly red colour resembles the complex nature of the objects in the LRNe group, with different scaling. However, such behaviour is also shown by objects interpreted as a SN impostor, such as SN~Hunt~248, in NGC 5806 \citep{Mauerhan2015,Kankare2015}. 

The lack of periodic microvariation in the lightcurve $-$15 to $-$5 years before outburst suggests that, unlike in the case of Galactic merger V1309 Sco, where both binary components were detected, for M101-OT only the brightest star in the system was seen. The unusual location of the progenitor in the Hertzsprung gap supports the hypothesis that star is quickly expanding after finishing the core H burning phase. If such star has a close companion, whenever it expands enough to overfill its Roche lobe, it will initiate the mass transfer towards the secondary, forming a CE surrounding the binary system, so that the accretor will become engulfed in the envelope of the donor star. 

Given the low densities in the outer layers of the donor atmosphere, the initial drag on the secondary may not be noticeable on short timescales. However, the spiral-in phase will accelerate with the secondary orbiting in increasingly denser layers of the primary star, eventually leading either to the merger of the components or the ejection of the envelope of the primary star on dynamical timescales. The slow brightening in M101-OT before the detected outbursts could have been associated with these final stages. The existence of optically thick ejected material is confirmed by the quick colour evolution of M101-OT in the blue bands, which suggests the existence optically thick ejected material. 

The spectrum of M101-OT is dominated by H$\alpha$, Ca II, Ba II, Na II and K I at low expansion velocities ($\sim 300$ \kms) and a forest of Ti II and Fe II absorption lines at short wavelengths. These characteristics are similar to other LRNe, such as V838 Mon, M31 LRN or NGC 4490 2011OT-1. However, low expansion velocities are not exclusive of this class. Members of LBVs and ILOT classes also show outflow velocities well below 1000 \kms. The double-peaked H$\alpha$ emission profile, tracing the bipolar structure of the ejecta, has also been observed in the asymmetric outflows of LBVs \citep{Smith2016b} and nebular phases of SNe IIn with bipolar CSM \citep{Smith2015,Andrews2016} or CCSNe, such as SN 1987A \citep{Groningsson2008}. Newly formed dust within the ejecta is the responsible for the extinction of optical and NIR light. The redshifted component undergoes greater absorption from the generated dust, and therefore the blue emission may become more dominant at late epochs \citep{Bevan2016}. 

One distinctive feature of M101-OT is the prompt formation of molecular bands, which suggest the presence of newly formed dust. At +154 days the spectrum showed evidence of formation of TiO and VO bands, comparable to the ones seen in LRNe V4332 Sgr \citep{Martini1999,Kaminski2010} and V838 Mon \cite{Rushton2005,Tylenda2011}.  Figure \ref{fig:m5iii} shows a comparison of the spectrum of M101-OT at +154 days, along with the UVES/VLT average spectrum of V838 Mon taken in January, February and March 2009, about seven years after the outburst \citep{Tylenda2011}. Although the resolution of the GTC spectrum is not high enough (380 \kms) to resolve individual bands, they match well with a spectrum of an M5III star \citep{Bagnulo2003}. 

Possible interpretations of the true nature of M101-OT may include a wide range of scenarios. Some examples are: onset of the CE, similar to the one witnessed for M31 2015 LRN \citep{MacLeod2016}; mass-loss during turbulent phases of the stellar evolution (e.g. during the post He-burning phase); mass loss events triggered by the passage of a lower-mass companion to the periastron and the subsequent shell-shell collision in very eccentric orbits; swallowing of planets by an expanding red giant \citep{Retter2003}; mass transfer induced jets, similar to the ones suggested for M31 2015 LRN and SN~2015bh \citep{SokerKashi2016,Soker2016}; a faint terminal explosion or even thermal emission from shocks originated from the mass loss in the binary system \citep{Pejcha2016}. The binary merger scenario has also been proposed by \cite{Goranskij2016}, who interpreted M101-OT as a massive OB binary system.

We argue that, within the context of binary evolution models, M101-OT likely represents the best studied case of an unusual event of the ejection of the CE in a massive binary system, leaving two surviving components in a closer massive binary system formed by the lower mass main sequence star and the He core of the stripped companion. The characteristics of the M101-OT agree with the empirical correlation between the peak absolute magnitude in $I$-band and the progenitor mass suggested by \cite{Kochanek2014}. Future surveys targeting nearby galaxies would help to populate this correlation in the more massive end.

\section{Summary and conclusions}\label{sec:conclusions}

M101-OT is a transient with LRN characteristics discovered in a star forming region in a spiral arm of M101. A summary of its most relevant observational characteristics is given below:
\begin{itemize}
\renewcommand\labelitemi{--}
\item The historic evolution of M101-OT shows no major variations within 0.2 mag in $R$-band until approximately 5.5 years before the outburst.
 \item The pre-outburst SED suggests no IR excess, implying the lack of an old existing dust emission component.
\item The object has slowly brightened by 1.5 mags over the last 6 years prior to the outburst. The estimated radius appeared to increase from 230$~\pm~13$ \Rsun at 6 years before the outburst, to 3100 \Rsun during the secondary outburst maximum.
\item The lightcurve shows two peaks, detected in $R$-band, separated by $\geq$ 100 days. The magnitude of the first peak is $M_r\ \leq\ -12.4$ mag (lower limit because of an observation gap) and  $M_r\ \simeq\ -12.0$ during the second peak. The colour of the object during the second maximum is $g-r=1.4$ mag, which corresponds to an estimated temperature of 3600 K.
\item Late time follow-up photometry suggests the re-brightening of the object in IR wavelengths after one year.
\item The bolometric luminosity for the second peak is $L~=~2.7\times\ 10^{40}$ ergs s$^{-1}$ and the total energy release during the outburst is $L\ > \ 4.1 \times 10^{47}$ erg. This is only a lower limit, as the first outburst is not covered well enough to put a tight constraint on the energy.
\item During peak, the spectrum shows a cold photospheric continuum, combined with low expansion velocities ($\sim 300$ \kms) for H$\alpha$, Fe II and low energy ionization elements, which display a P-Cygni profile.
\item The lightcurve after the second outburst is defined by a short decline phase ($\sim$ 40 days), a ``plateau'' phase ($\sim$60 days) in $riz$ bands and a second decline phase. The photospheric radius at the beginning of each phase was $\sim$ 6500 \Rsun, 4300 \Rsun and 5800 \Rsun respectively.
 \item The H$\alpha$ line shows initially a blueshifted absorption component at $-$500 \kms, which develops into an emission profile at epochs +30 days or later.
\item The spectrum shows the formation of molecular bands after 100 days of the outburst, which suggests the fast formation of dust in the system.
\item The best fit for the progenitor is an F-type giant with a luminosity of L$\sim8.7\ \times\ 10^4$ \Lsun and initial mass of 18$\pm$1. The estimated age of the star is 9.89, which places it in the Hertzsprung-Russell gap. The age is qualitatively consistent with the young stellar population surrounding the progenitor, although high accuracy photometry will be needed to provide a quantitative answer.
\item In the binary case scenario, assuming that the primary is overfilling its Roche lobe, the binary system is initially on a wide orbit, with periods between 600 and 270 days (for $q=1$ and $q=18$ respectively). By the end of the common envelope phase, the fate of the system depends on the model. While the simple energy formalism anticipates the complete merger of the system, binary evolution models favor the survival of the binary stellar component with 260$-$290 day period.
\end{itemize}

Although the nature of the object is not entirely clear, its resemblance with other transients from the same LRN family points towards a possible binary origin. The unusual location of the progenitor star in the Hertzsprung gap supports the hypothesis that the most massive component had expanded beyond its Roche lobe, initiating the CE phase. The outbursts detected for M101-OT suggest that this CE was ejected on dynamical timescales, likely leaving a surviving close binary pair. 

We have discussed the past and present evolution of this unusual transient in M101; discussion of its future and the fate of its remnant will have to await further observations in the IR bands.

\section*{Acknowledgments} \label{sec:acknowledgements}

\small
The research leading to these results has received funding from the European Union Seventh Framework Programme ([FP7/2007-2013] under grant agreement num. 264895. This work was partly supported by the European Union FP7 programme through ERC grant number 320360. This work was supported, in whole or in part, by the European Science Foundation under the GREAT ESF RNP programme. This work was supported by the GROWTH project funded by the National Science Foundation under Grant No 1545949.
LANL participation in iPTF was funded by the US Department of Energy as
part of the Laboratory Directed Research and Development program.
Part of this research was carried out at the Jet Propulsion Laboratory,
California Institute of Technology, under a contract with the National
Aeronautics and Space Administration.
Based on observations obtained with MegaPrime/MegaCam, a joint project of CFHT and CEA/IRFU, at the Canada-France-Hawaii Telescope (CFHT) which is operated by the National Research Council (NRC) of Canada, the Institut National des Science de l'Univers of the Centre National de la Recherche Scientifique (CNRS) of France, and the University of Hawaii. This work is based in part on data products produced at Terapix available at the Canadian Astronomy Data Centre as part of the Canada-France-Hawaii Telescope Legacy Survey, a collaborative project of NRC and CNRS. This paper makes use of data obtained from the Isaac Newton Group Archive which is maintained as part of the CASU Astronomical Data Centre at the Institute of Astronomy, Cambridge. This work is partly based on observations obtained with the Nordic Optical Telescope, operated by the Nordic Optical Telescope Scientific Association at the Observatorio del Roque de los Muchachos, La Palma, Spain. This work is partly based on observations made with the William Hershell Telescope operated on the island of La Palma by the Isaac Newton Group in the Spanish Observatorio del Roque de los Muchachos of the Instituto de Astrofísica de Canarias. The Gran Telescopio Canarias (GTC) operated on the island of La Palma at the Spanish Observatorio del Roque de los Muchachos of the Instituto de Astrofisica de Canarias. This work is partly based on data from Copernico 1.82m telescope operated by INAF Osservatorio Astronomico di Padova. NER, AP, GT and MT are partially supported by the PRIN-INAF 2014 with the project ``Transient Universe: unveiling new types of stellar explosions with PESSTO''. Finally, NBM would like to thank Robert G. Izzard, Philipp Podsiadlowski, Lars Bildsten, E. Sterl Phinney and Noam Soker for helpful discussions, and Pablo and Lucia Solis, and Israel Zenteno for the motivation.

\facility{Asiago:Copernico, CFHT, GTC, Hale, ING:Newton, ING:Herschel, Keck:I, LBT, Liverpool:2m, NOT, PO:1.2m, PO:1.5m, PS1, Sloan, Spitzer, UKIRT }

\software{BPASS v1.0, BPASS v2.0, EMCEE \citep{Foreman-Mackey2013}, IRAF, NOTCAM (v2.54), PTFIDE \citep{Masci2016}, PYRAF, SNOoPY}.

\appendix\label{appendix}
\section{Photometry tables}


\onecolumngrid
\newpage 

\renewcommand{\tabcolsep}{0.02cm}

\begin{deluxetable*}{rccccccccccccccc} 
\tabletypesize{\scriptsize} 
\tablecolumns{10} 
\tablecaption{ Historic photometric measurements of M101-OT.\label{table:preoutburst}} 
\tablehead{ 
\colhead{ Phase}  & \colhead{ MJD } & \colhead{ Tel.} & 
\colhead{ $m_U$ }& \colhead{ $m_B$} & \colhead{ $m_V$ } 
& \colhead{ $m_R$ }  & \colhead{ $m_u$ } & \colhead{ $m_g$ }  & 
\colhead{ $m_r$ }  & \colhead{ $m_i$ }  & \colhead{ $m_z$} &
 \colhead{ $m_y$ } & \colhead{ Unfilt. } & \colhead{ Refs.}\\ 
\colhead{ (days) } 
& \colhead{  (+50000)    } 
& \colhead{   } & \colhead{ (mag)} & \colhead{ (mag) } & \colhead{ (mag) } & \colhead{ (mag) } & \colhead{ (mag) } & \colhead{ (mag) } & \colhead{ (mag) } & \colhead{ (mag) } & \colhead{ (mag)} & \colhead{ (mag) } & \colhead{ (mag) } & \colhead{ }\\ 
} 
\startdata 
 $-$5413.5 & 1656.5 & CFHT & -- & 21.94$\pm$0.11 & 21.89$\pm$0.31 & 21.16$\pm$0.19 & -- & -- & -- & -- & -- & -- & --&\\
 $-$5410.8 & 1659.2 & INT & -- & -- & -- & -- & -- & -- & 21.17$\pm$0.17 & -- & -- & -- & --&\\
 $-$4364.0 & 2706.0 & SDSS & -- & -- & -- & -- & 23.13$\pm$0.49 & 21.79$\pm$0.09 & 21.86$\pm$0.14 & 21.31$\pm$0.08 & 21.94$\pm$0.58 & -- & --& [1]\\
 $-$3920.1 & 3149.9 & INT & -- & 21.83$\pm$0.22 & -- & -- & -- & -- & -- & -- & -- & -- & --&\\
 $-$3860.6 & 3209.4 & CFHT & -- & -- & -- & -- & 21.82$\pm$0.36 & 21.38$\pm$0.13 & -- & -- & -- & -- & --&\\
  ... & ... & ... & ... & ... & ... & ... & ... & ... & ... & ... & ... & ... & ... & ... \\
\enddata 
\tablecomments{References: [1], \cite{ATel7082}. Table \ref{table:preoutburst} is published in its entirety in the machine-readable format. A portion is shown here for guidance regarding its form and content.} 
\end{deluxetable*}

\begin{deluxetable*}{rccccccccccccc} 
\tabletypesize{\scriptsize} 
\tablecolumns{14} 
\tablecaption{ Post-discovery photometric measurements of M101-OT. \label{table:followup}} 
\tablehead{ 
\colhead{ Phase}  & \colhead{ MJD } & \colhead{ Tel. } & 
\colhead{ $m_U$ }& \colhead{ $m_B$} & \colhead{ $m_V$ } 
& \colhead{ $m_R$ } & \colhead{ $m_I$ }  & \colhead{ $m_u$ } & \colhead{ $m_g$ }  & 
\colhead{ $m_r$ }  & \colhead{ $m_i$ }  & \colhead{ $m_z$}  & \colhead{ Refs.}\\ 
\colhead{ (days) } 
& \colhead{  (+50000)    } 
& \colhead{   } & \colhead{ (mag)} & \colhead{ (mag) } & \colhead{ (mag) } & \colhead{ (mag) } & \colhead{ (mag) } & \colhead{ (mag) } & \colhead{ (mag) } & \colhead{ (mag) } & \colhead{ (mag)} & \colhead{ (mag) } & \colhead{ }\\ 
} 
\startdata 
 4.0 & 7074.0 & SAI$-$2.5m & -- & 19.09$\pm$0.02 & 17.71$\pm$0.02 & 16.83$\pm$0.02 & -- & -- & -- & -- & -- & --& [1]\\
 7.9 & 7077.9 & SAO$-$6m & -- & 19.04$\pm$0.02 & 17.69$\pm$0.02 & 16.83$\pm$0.02 & -- & -- & -- & -- & -- & --& [1]\\
 11.1 & 7081.1 & SAI$-$2.5m & -- & 19.23$\pm$0.02 & 17.79$\pm$0.02 & 16.90$\pm$0.02 & -- & -- & -- & -- & -- & --& [1]\\
 17.1 & 7087.1 & NOT & -- & 19.57$\pm$0.05 & 18.12$\pm$0.02 & 17.07$\pm$0.01 & 16.35$\pm$0.05 & -- & -- & -- & -- & --&\\
 22.2 & 7092.2 & LT & -- & 19.90$\pm$0.26 & -- & -- & -- & -- & 19.06$\pm$0.20 & 17.73$\pm$0.06 & 17.23$\pm$0.08 & 16.88$\pm$0.07&\\
 ... & ... & ... & ... & ... & ... & ... & ... & ... & ... & ... & ... & ... &... \\
\enddata 
\tablecomments{References: [1] \cite{ATel7206}. Table \ref{table:followup} is published in its entirety in the machine-readable format. A portion is shown here for guidance regarding its form and content.} 
\end{deluxetable*}

\begin{deluxetable}{rccccc} 
\tabletypesize{\tiny} 
\tablecolumns{6} 
\tablecaption{ iPTF follow-up data of M101-OT.\label{table:ptf}} 
\tablehead{ 
\colhead{ Phase}  & \colhead{ MJD } & \colhead{ Telescope } & 
 \colhead{ $m_g$ }  & 
\colhead{ $m_r$ }  & \colhead{ $m_i$ }  \\ 
 \colhead{ (days) } 
& \colhead{  (+50000)    } 
& \colhead{   } & \colhead{ (mag)} & \colhead{ (mag) } & \colhead{ (mag) } \\ 
} 
\startdata 
$-$2005.8 & 5064.2 & PTFP48 & -- & 20.26$\pm$0.11 & --\\
 $-$1706.7 & 5363.3 & PTFP48 & -- & 20.31$\pm$0.11 & --\\
 $-$1682.7 & 5387.3 & PTFP48 & -- & 20.18$\pm$0.15 & --\\
 $-$1484.6 & 5585.4 & PTFP48 & -- & 20.17$\pm$0.13 & --\\
 $-$1345.7 & 5724.3 & PTFP48 & -- & 20.25$\pm$0.14 & --\\ 
... & ... & ... & ... & ... & ...\\
\enddata 
\tablecomments{The errors are given in brackets. Table \ref{table:ptf} is published in its entirety in the machine-readable format. A portion is shown here for guidance regarding its form and content.} 
\end{deluxetable}

\begin{table*}
\centering
\caption{Sequence star magnitudes used for the M101 field. The values are computed from the stacked magnitudes from Pan-STARRS.}
\begin{small}
\centering
\begin{tabular}{ccccccccc}
\hline
\hline
Star & $\alpha$ & $\delta$ & $m_{g}$ & $m_{r}$ & $m_{i}$ & $m_{z}$ & $m_{y}$ \\
\# & (J2000.0) & (J2000.0) & (mag) & (mag) & (mag) & (mag) \\ 
\hline
1 & 210.6328 & 54.4887 & 18.347 (0.007) & 17.488 (0.004)& 17.123 (0.004)& 16.935 (0.006)& 16.836 (0.007)\\
2 & 210.5005 & 54.4851 & 17.463 (0.005) & 16.388 (0.005)& 15.776 (0.004)& 15.498 (0.004)& 15.357 (0.004)\\
3 & 210.6519 & 54.4464 & 16.426 (0.004) & 15.692 (0.004)& 15.398 (0.004)& 15.277 (0.004)& 15.190 (0.005)\\
4 & 210.6370 & 54.4479 & 16.863 (0.006) & 15.915 (0.005)& 15.534 (0.005)& 15.359 (0.003)& 15.251 (0.006)\\
5 & 210.5874 & 54.4614 & 17.171 (0.007) & 16.394 (0.005)& 16.097 (0.005)& 15.954 (0.003)& 15.872 (0.005)\\
6 & 210.5708 & 54.4583 & 17.300 (0.007) & 16.263 (0.004)& 15.806 (0.004)& 15.580 (0.004)& 15.457 (0.005)\\
7 & 210.5798 & 54.4493 & 15.396 (0.005) & 14.510 (0.004)& 14.128 (0.004)& 13.940 (0.003)& 13.830 (0.004)\\
8 & 210.6340 & 54.4175 & 19.256 (0.012) & 18.031 (0.006)& 17.273 (0.004)& 16.915 (0.005)& 16.745 (0.008)\\
9 & 210.6171 & 54.4159 & 17.990 (0.011) & 16.725 (0.004)& 15.850 (0.005)& 15.444 (0.003)& 15.242 (0.005)\\
10 & 210.6189 & 54.4048 & 16.238 (0.004) & 15.794 (0.003)& 15.631 (0.003)& 15.577 (0.003)& 15.547 (0.005)\\
11 & 210.4896 & 54.4296 & 16.789 (0.011) & 16.347 (0.004)& 16.250 (0.004)& 16.183 (0.004)& 16.172 (0.007)\\
12 & 210.5186 & 54.4001 & 18.317 (0.007) & 17.516 (0.004)& 17.191 (0.005)& 17.012 (0.006)& 16.901 (0.007)\\
13 & 210.5614 & 54.4480 & 21.791 (0.110) & 21.516 (0.062)& 21.341 (0.060)& 20.938 (0.205)& -- (--)\\
14 & 210.5582 & 54.4429 & 19.752 (0.022) & 19.292 (0.010)& 19.133 (0.010)& 19.080 (0.017)& 18.984 (0.039)\\
15 & 210.5892 & 54.4280 & 19.388 (0.012) & 18.499 (0.007)& 18.187 (0.005)& 18.024 (0.008)& 17.918 (0.018)\\
16 & 210.5873 & 54.4296 & 22.027 (0.185) & 21.018 (0.051)& 19.627 (0.013)& 19.041 (0.018)& 18.646 (0.028)\\
17 & 210.5272 & 54.4393 & 21.070 (0.036) & 20.555 (0.028)& 20.313 (0.022)& 20.209 (0.041)& 19.641 (0.107)\\
18 & 210.5204 & 54.4523 & 21.392 (0.048) & 21.017 (0.039)& 20.799 (0.034)& 20.758 (0.072)& -- (--)\\
19 & 210.4864 & 54.4320 & 19.935 (0.017) & 18.712 (0.007)& 17.673 (0.004)& 17.199 (0.005)& 16.962 (0.010)\\
\hline
\end{tabular}
\end{small}
\begin{tablenotes}
    \item[\textdagger] The errors are given in brackets. Coordinates and magnitudes are taken from the Pan-STARRS PV2 Catalog.
    \end{tablenotes}
    \label{table:seq}
\end{table*}

\renewcommand{\tabcolsep}{0.3cm}
\begin{deluxetable*}{rccccccc} 
\tabletypesize{\tiny} 
\tablecolumns{8} 
\tablecaption{NIR and MIR photometry of M101-OT.\label{table:irphot}} 
\tablehead{ 
\colhead{ Phase}  & \colhead{ MJD } & \colhead{ Telescope } & 
 \colhead{ $J$ }  & \colhead{ $H$ }  & \colhead{ $K$ }  & \colhead{ $3.6\mu $m} & \colhead{$4.5 \mu $m}  \\ 
\colhead{ (days) } & \colhead{  (+50000)} & \colhead{   } & \colhead{ (mag)} & \colhead{ (mag) } & \colhead{ (mag) }  & \colhead{ (mag) }  & \colhead{ (mag) } \\ 
} 
\startdata 
 -4065.0 & 3005.0 & Spitzer & -- & -- & --& 19.00$\pm$0.01& 18.60$\pm$0.01\\
 -3997.5 & 3072.5 & Spitzer & -- & -- & -- & --& $>$20.48\\
 -2075.0 & 4995.0 & CFHT& 22.74$\pm$0.53 & --& 19.54$\pm$0.11 & -- & --\\
 -1756.6 & 5313.4 & UKIRT& 20.61$\pm$0.16 & -- & -- & -- & --\\
 17.1 & 7087.1 & NOT& 15.48$\pm$0.03& 15.10$\pm$0.06& 14.97$\pm$0.09 & -- & --\\
 48.0 & 7118.0 & NOT& 16.42$\pm$0.03& 15.74$\pm$0.05& 15.47$\pm$0.01 & -- & --\\
 62.6 & 7132.6 & Spitzer & -- & -- & --& 14.72$\pm$0.01& 14.75$\pm$0.01\\
 74.0 & 7144.0 & NOT& 16.70$\pm$0.04& 15.72$\pm$0.02& 15.49$\pm$0.07 & -- & --\\
 90.3 & 7160.3 & Spitzer & -- & -- & --& 14.55$\pm$0.01& 14.59$\pm$0.01\\
 102.0 & 7172.0 & NOT& 16.63$\pm$0.03& 15.90$\pm$0.04& 15.41$\pm$0.06 & -- & --\\
 121.8 & 7191.8 & Spitzer & -- & -- & --& 14.50$\pm$0.01 & --\\
 136.0 & 7206.0 & NOT& 16.53$\pm$0.02& 16.01$\pm$0.03& 15.46$\pm$0.04 & -- & --\\
 150.8 & 7220.8 & Spitzer & -- & -- & --& 14.39$\pm$0.01 & --\\
 197.9 & 7267.9 & NOT& 17.16$\pm$0.03& 16.02$\pm$0.03& 15.25$\pm$0.02 & -- & --\\
 202.7 & 7272.7 & Spitzer & -- & -- & --& 14.30$\pm$0.01& 14.50$\pm$0.01\\
 253.3 & 7323.3 & NOT & --& 17.87$\pm$0.07& 16.32$\pm$0.03 & -- & --\\
 280.3 & 7350.3 & NOT & --& $>$16.56& 16.81$\pm$0.07 & -- & --\\
 339.0 & 7409.0 & SAI-2.5m& 18.65$\pm$0.05& 17.64$\pm$0.10& 16.32$\pm$0.05 & -- & --\\
 385.0 & 7455.0 & SAI-2.5m& 21.00$\pm$0.20& 20.20$\pm$0.30& 18.20$\pm$0.20 & -- & --\\
 403.4 & 7473.4 & P200& 21.85$\pm$0.42& 19.59$\pm$0.17& 17.91$\pm$0.12 & -- & --\\
 438.0 & 7508.0 & NOT & -- & --& 17.22$\pm$0.12 & -- & --\\
 514.0 & 7584.0 & NOT& 20.01$\pm$0.24& 18.77$\pm$0.13& 17.25$\pm$0.13 & -- & --\\
\enddata 
\end{deluxetable*}

\renewcommand{\tabcolsep}{0.1cm}

\begin{table*}
\begin{minipage}{1.\linewidth}
\begin{small}
\caption{Log of spectroscopic observations of M101-OT.}
\centering
\begin{tabular}{rllccccc}
\hline
 Phase$^a$ & MJD & Date     & Telescope+Instrument & Grating$/$Grism & Dispersion & Resolution$^b$ & Exposure   \\ 
   (d) &  (+50000)	&  			&  					   &	& (\angstrom/pix)		& (km/s) &	(s)	     \\ \hline
 \hline
-2.5 & 7067.5 & 2015 Feb 14.5 & P200+DBSP & 7500 & 1.52 & 500 & 900  \\
-2.0 & 7068.0 & 2015 Feb 15 & NOT+ALFOSC & Grism\#4 & 3.0 & 733 & 1800  \\
0.1 & 7070.1 & 2015 Feb 17.1 & Copernico 1.82m+AFOSC & GR04 & 4.3 & 630 & 1800  \\
1.5 & 7071.5 & 2015 Feb 18.5 & P200+DBSP & 158/7500 &  1.52 & 420 & 1800  \\
2.1 & 7072.1 & 2015 Feb 19.1 & Copernico 1.82m+AFOSC & GR04 & 4.3 & 690 & 1800 \\
22.9 & 7092.9 & 2015 Mar 11.9 & WHT+ISIS			& R300B+R158R & 0.86+1.8 & 320 & 2$\times$600+2$\times$600  \\
54.2 & 7124.2 & 2015 Apr 13.2 & NOT+ALFOSC		& Grism\#4 & 3.0 & 747 & 2700 \\
116.1 & 7186.3 & 2015 Jun 13.3 & WHT+ISIS 			& R158R & 1.8 & 345 & 2$\times$1800  \\
153.9 & 7223.9 & 2015 Jul 20.9 & GTC+OSIRIS & R1000B & 2.1 &380& 2$\times$1800.0  \\
 \hline
\end{tabular}
\begin{tablenotes}
    \item[\textdagger]$^a$ Days since second peak at MJD 57070. $^b$ Measured using the FWHM of $\lambda$ 5577 O I sky line.
    \end{tablenotes}
\label{table:speclog}
\end{small}
\end{minipage}
\end{table*}

\label{lastpage}

\afterpage{}
\clearpage

\bibliographystyle{aasjournal}
\bibliography{main}

\end{document}


\renewcommand{\tabcolsep}{0.05cm}

\begin{deluxetable}{rccccccccccccccp{1cm}} 
\tabletypesize{\scriptsize} 
\tablewidth{\textwidth} 
\tablecaption{ Historic photometric measurements of M101-OT.\label{table:preoutburst}} 
\tablehead{ 
\colhead{ Phase}  & \colhead{ MJD } & \colhead{ Telescope } & 
\colhead{ $m_U$ }& \colhead{ $m_B$} & \colhead{ $m_V$ } 
& \colhead{ $m_R$ } & \colhead{ $m_u$ } & \colhead{ $m_g$ }  & 
\colhead{ $m_r$ }  & \colhead{ $m_i$ }  & \colhead{ $m_z$} &
 \colhead{ $m_y$ } & \colhead{ Unfilt. } & \colhead{ Refs.}\\ 
\colhead{ (days) } 
& \colhead{  (+50000)    } 
& \colhead{   } & \colhead{ (mag)} & \colhead{ (mag) } & \colhead{ (mag) } & \colhead{ (mag) } & \colhead{ (mag) } & \colhead{ (mag) } & \colhead{ (mag) } & \colhead{ (mag) } & \colhead{ (mag)} & \colhead{ (mag) } & \colhead{ (mag) } & \colhead{ }\\ 
} 
\startdata 
  $-$5413.5 & 1656.5 & CFHT & -- & 21.91$\pm$0.11 & 21.87$\pm$0.31 & 21.14$\pm$0.19 & -- & -- & -- & -- & -- & -- & --&\\
 $-$5410.8 & 1659.2 & INT & -- & -- & -- & -- & -- & -- & 21.15$\pm$0.17 & -- & -- & -- & --&\\
 $-$4364.0 & 2706.0 & SDSS & -- & -- & -- & -- & 23.09$\pm$0.49 & 21.76$\pm$0.09 & 21.84$\pm$0.14 & 21.29$\pm$0.08 & 21.93$\pm$0.58 & -- & --& [1]\\
 $-$3920.1 & 3149.9 & INT & -- & 21.80$\pm$0.22 & -- & -- & -- & -- & -- & -- & -- & -- & --&\\
 $-$3860.6 & 3209.4 & CFHT & -- & -- & -- & -- & 21.78$\pm$0.36 & 21.36$\pm$0.13 & -- & -- & -- & -- & --&\\
 $-$3652.4 & 3417.6 & CFHT & -- & -- & -- & -- & -- & -- & 21.19$\pm$0.06 & -- & -- & -- & --&\\
 $-$3598.5 & 3471.5 & CFHT & -- & -- & -- & -- & -- & -- & 21.09$\pm$0.09 & -- & -- & -- & --&\\
 $-$3569.6 & 3500.4 & CFHT & -- & -- & -- & -- & -- & 21.70$\pm$0.10 & 21.31$\pm$0.08 & -- & -- & -- & --&\\
 $-$3569.0 & 3501.0 & CFHT & -- & -- & -- & -- & -- & -- & -- & 21.45$\pm$0.11 & -- & -- & --&\\
 $-$3566.6 & 3503.4 & CFHT & -- & -- & -- & -- & -- & 21.56$\pm$0.07 & -- & -- & -- & -- & --&\\
 $-$3565.7 & 3504.3 & CFHT & -- & -- & -- & -- & -- & -- & -- & 21.40$\pm$0.06 & -- & -- & --&\\
 $-$3563.6 & 3506.4 & CFHT & -- & -- & -- & -- & -- & -- & 21.02$\pm$0.05 & -- & -- & -- & --&\\
 $-$3299.4 & 3770.6 & CFHT & -- & -- & -- & -- & -- & 21.36$\pm$0.12 & 21.16$\pm$0.21 & -- & -- & -- & --&\\
 $-$3298.4 & 3771.6 & CFHT & -- & -- & -- & -- & -- & -- & -- & 21.30$\pm$0.25 & 21.69$\pm$0.56 & -- & --&\\
 $-$3178.6 & 3891.4 & CFHT & -- & -- & -- & -- & -- & -- & -- & -- & 21.32$\pm$0.07 & -- & --&\\
 $-$3177.0 & 3893.0 & CFHT & -- & -- & -- & -- & -- & -- & -- & -- & 21.22$\pm$0.04 & -- & --&\\
 $-$2866.6 & 4203.4 & CFHT & -- & -- & -- & -- & 21.75$\pm$0.13 & -- & -- & -- & -- & -- & --&\\
 $-$2557.4 & 4512.6 & CFHT & -- & -- & -- & -- & -- & -- & 21.06$\pm$0.04 & -- & -- & -- & --&\\
 $-$2535.0 & 4535.0 & LBT & 21.29$\pm$0.19 & 21.27$\pm$0.19 & 20.95$\pm$0.17 & 20.67$\pm$0.17 & -- & -- & -- & -- & -- & -- & --& [2]\\
 $-$2450.7 & 4619.3 & CFHT & -- & -- & -- & -- & 21.90$\pm$0.37 & -- & -- & -- & -- & -- & --&\\
 $-$1838.4 & 5231.6 & PS1 & -- & -- & -- & -- & -- & -- & -- & -- & 20.94$\pm$0.22 & -- & --&\\
 $-$1485.3 & 5584.7 & PS1 & -- & -- & -- & -- & -- & -- & -- & -- & 20.71$\pm$0.05 & -- & --&\\
 $-$1398.6 & 5671.4 & PS1 & -- & -- & -- & -- & -- & -- & -- & 21.00$\pm$0.13 & -- & -- & --&\\
 $-$1396.6 & 5673.4 & PS1 & -- & -- & -- & -- & -- & -- & 20.69$\pm$0.12 & 20.67$\pm$0.12 & -- & -- & --&\\
 $-$1393.6 & 5676.4 & PS1 & -- & -- & -- & -- & -- & 21.01$\pm$0.32 & -- & -- & -- & -- & --&\\
 $-$1392.6 & 5677.4 & PS1 & -- & -- & -- & -- & -- & 21.01$\pm$0.16 & -- & -- & -- & -- & --&\\
 $-$1377.7 & 5692.3 & PS1 & -- & -- & -- & -- & -- & -- & 20.32$\pm$0.12 & -- & -- & -- & --&\\
 $-$1098.3 & 5971.7 & PS1 & -- & -- & -- & -- & -- & -- & -- & -- & 20.49$\pm$0.20 & -- & --&\\
 $-$1040.5 & 6029.5 & PS1 & -- & -- & -- & -- & -- & -- & 20.56$\pm$0.12 & -- & -- & -- & --&\\
 $-$1040.5 & 6029.5 & PS1 & -- & -- & -- & -- & -- & -- & -- & 20.72$\pm$0.12 & -- & -- & --&\\
 $-$1038.5 & 6031.5 & PS1 & -- & -- & -- & -- & -- & 20.82$\pm$0.09 & -- & -- & -- & -- & --&\\
 $-$1014.6 & 6055.4 & PS1 & -- & -- & -- & -- & -- & -- & -- & -- & 20.60$\pm$0.22 & -- & --&\\
 $-$689.4 & 6380.6 & PS1 & -- & -- & -- & -- & -- & -- & -- & 20.46$\pm$0.47 & -- & -- & --&\\
 $-$681.6 & 6388.4 & PS1 & -- & -- & -- & -- & -- & 20.62$\pm$0.17 & 20.37$\pm$0.19 & -- & -- & -- & --&\\
 $-$674.6 & 6395.4 & PS1 & -- & -- & -- & -- & -- & 20.47$\pm$0.18 & -- & -- & -- & -- & --&\\
 $-$664.6 & 6405.4 & PS1 & -- & -- & -- & -- & -- & -- & 20.11$\pm$0.05 & -- & -- & -- & --&\\
 $-$633.6 & 6436.4 & PS1 & -- & -- & -- & -- & -- & -- & -- & -- & 20.23$\pm$0.19 & 20.39$\pm$0.11 & --&\\
 $-$358.4 & 6711.6 & PS1 & -- & -- & -- & -- & -- & -- & 19.91$\pm$0.23 & -- & -- & -- & --&\\
 $-$334.5 & 6735.5 & PS1 & -- & -- & -- & -- & -- & -- & 19.76$\pm$0.05 & -- & 19.87$\pm$0.11 & -- & --&\\
 $-$284.7 & 6785.3 & PS1 & -- & -- & -- & -- & -- & -- & -- & 19.78$\pm$0.08 & -- & -- & --&\\
 $-$250.0 & 6820.0 & LBT & 20.39$\pm$0.02 & 19.99$\pm$0.04 & 19.76$\pm$0.03 & 19.57$\pm$0.03 & -- & -- & -- & -- & -- & -- & --& [2]\\
 $-$227.7 & 6842.3 & PS1 & -- & -- & -- & -- & -- & -- & -- & 19.71$\pm$0.06 & -- & -- & --&\\
 $-$205.8 & 6864.2 & PS1 & -- & -- & -- & -- & -- & -- & -- & -- & -- & 19.52$\pm$0.21 & --&\\
 $-$96.0 & 6974.0 & Amateurs & -- & -- & 16.38$\pm$0.30 & -- & -- & -- & -- & -- & -- & -- & --&\\
 $-$58.0 & 7012.0 & Amateurs & -- & -- & -- & -- & -- & -- & -- & -- & -- & -- & 17.90$\pm$0.30&\\
 $-$29.0 & 7041.0 & LBT & 21.00$\pm$0.02 & 20.17$\pm$0.02 & 18.78$\pm$0.02 & 18.21$\pm$0.02 & -- & -- & -- & -- & -- & -- & --& [2]\\
 $-$28.0 & 7042.0 & Amateurs & -- & -- & 18.48$\pm$0.30 & -- & -- & -- & -- & -- & -- & -- & --&\\
 $-$6.0 & 7064.0 & ASASSN & -- & -- & 17.22$\pm$0.00 & -- & -- & -- & -- & -- & -- & -- & --& [3]\\
 $-$6.0 & 7064.0 & Amateurs & -- & -- & -- & -- & -- & -- & -- & -- & -- & -- & 16.50$\pm$0.30&\\
 $-$5.0 & 7065.0 & ASASSN & -- & -- & 17.24$\pm$0.00 & -- & -- & -- & -- & -- & -- & -- & --& [3]\\
 $-$4.0 & 7066.0 & Amateurs & -- & -- & -- & -- & -- & -- & -- & -- & -- & -- & 16.70$\pm$0.30&\\
 $-$3.0 & 7067.0 & Amateurs & -- & -- & -- & -- & -- & -- & -- & -- & -- & -- & 17.10$\pm$0.30&\\
 $-$3.0 & 7067.0 & Swift & -- & 18.87$\pm$0.10 & 17.58$\pm$0.10 & -- & -- & -- & -- & -- & -- & -- & --& [4]\\
 $-$2.0 & 7068.0 & Amateurs & -- & -- & 17.48$\pm$0.30 & -- & -- & -- & -- & -- & -- & -- & --&\\
 $-$1.0 & 7069.0 & SAI$-$2.5m & -- & 18.83$\pm$0.02 & 17.65$\pm$0.02 & 16.77$\pm$0.02 & -- & -- & -- & -- & -- & -- & --& [5]\\
\enddata 
\tablecomments{References: 
[1] \cite{ATel7082}, 
[2] \cite{ATel7069}, 
[3] \cite{ATel7063}
[4] \cite{ATel7072}, 
[5] \cite{ATel7206}. } 
\end{deluxetable}

\begin{deluxetable*}{rccccccccccccc} 
\tabletypesize{\scriptsize} 
\tablecolumns{14} 
\tablecaption{ Post-discovery photometric measurements of M101-OT. \label{table:followup}} 
\tablehead{ 
\colhead{ Phase}  & \colhead{ MJD } & \colhead{ Telescope } & 
\colhead{ $m_U$ }& \colhead{ $m_B$} & \colhead{ $m_V$ } 
& \colhead{ $m_R$ } & \colhead{ $m_I$ }  & \colhead{ $m_u$ } & \colhead{ $m_g$ }  & 
\colhead{ $m_r$ }  & \colhead{ $m_i$ }  & \colhead{ $m_z$}  & \colhead{ Refs.}\\ 
\colhead{ (days) } 
& \colhead{  (+50000)    } 
& \colhead{   } & \colhead{ (mag)} & \colhead{ (mag) } & \colhead{ (mag) } & \colhead{ (mag) } & \colhead{ (mag) } & \colhead{ (mag) } & \colhead{ (mag) } & \colhead{ (mag) } & \colhead{ (mag)} & \colhead{ (mag) } & \colhead{ }\\ 
} 
\startdata 
 4.0 & 7074.0 & SAI$-$2.5m & -- & 19.06$\pm$0.02 & 17.69$\pm$0.02 & 16.81$\pm$0.02 & -- & -- & -- & -- & -- & --& [1]\\
 7.9 & 7077.9 & SAO$-$6m & -- & 19.01$\pm$0.02 & 17.67$\pm$0.02 & 16.81$\pm$0.02 & -- & -- & -- & -- & -- & --& [1]\\
 11.1 & 7081.1 & SAI$-$2.5m & -- & 19.20$\pm$0.02 & 17.77$\pm$0.02 & 16.88$\pm$0.02 & -- & -- & -- & -- & -- & --& [1]\\
 17.1 & 7087.1 & NOT & -- & 19.54$\pm$0.05 & 18.09$\pm$0.02 & 17.05$\pm$0.01 & 16.33$\pm$0.05 & -- & -- & -- & -- & --&\\
 22.2 & 7092.2 & LT & -- & 19.87$\pm$0.26 & -- & -- & -- & -- & 19.03$\pm$0.20 & 17.71$\pm$0.06 & 17.21$\pm$0.08 & 16.87$\pm$0.07&\\
 48.1 & 7118.1 & NOT & -- & 20.80$\pm$0.09 & 19.60$\pm$0.04 & 18.43$\pm$0.05 & 17.67$\pm$0.04 & -- & -- & -- & -- & --&\\
 48.1 & 7118.1 & LT & -- & 20.43$\pm$0.30 & -- & -- & -- & -- & 20.12$\pm$0.17 & 18.86$\pm$0.05 & 18.28$\pm$0.08 & 17.74$\pm$0.06&\\
 61.1 & 7131.1 & LT & -- & 21.23$\pm$0.16 & -- & -- & -- & -- & 20.30$\pm$0.06 & 18.99$\pm$0.03 & 18.36$\pm$0.11 & 17.87$\pm$0.03&\\
 69.1 & 7139.1 & LT & -- & 21.57$\pm$0.26 & -- & -- & -- & -- & 20.55$\pm$0.08 & 19.06$\pm$0.05 & 18.42$\pm$0.04 & 17.91$\pm$0.04&\\
 74.0 & 7144.0 & NOT & 19.91$\pm$0.21 & 21.41$\pm$0.14 & 20.07$\pm$0.04 & 18.70$\pm$0.04 & 17.88$\pm$0.03 & -- & -- & -- & -- & --&\\
 82.0 & 7152.0 & LT & -- & 21.88$\pm$0.35 & -- & -- & -- & -- & 20.69$\pm$0.15 & 19.21$\pm$0.03 & 18.48$\pm$0.06 & 17.90$\pm$0.03&\\
 90.1 & 7160.1 & LT & -- & 21.55$\pm$0.13 & -- & -- & -- & -- & 20.73$\pm$0.06 & 19.23$\pm$0.04 & 18.48$\pm$0.02 & 17.93$\pm$0.04&\\
 94.0 & 7164.0 & LT & -- & 21.89$\pm$0.33 & -- & -- & -- & -- & 20.89$\pm$0.13 & 19.26$\pm$0.06 & 18.48$\pm$0.09 & 17.91$\pm$0.04&\\
 96.9 & 7166.9 & LT & -- & 21.80$\pm$0.22 & -- & -- & -- & -- & 20.80$\pm$0.09 & 19.25$\pm$0.03 & 18.48$\pm$0.05 & 17.89$\pm$0.01&\\
 100.0 & 7170.0 & LT & -- & 22.28$\pm$0.36 & -- & -- & -- & -- & 20.79$\pm$0.12 & 19.31$\pm$0.05 & 18.50$\pm$0.05 & 17.94$\pm$0.06&\\
 102.1 & 7172.1 & NOT & -- & 21.65$\pm$0.00 & 19.95$\pm$0.08 & 19.03$\pm$0.06 & 17.68$\pm$0.04 & -- & -- & -- & -- & --&\\
 111.9 & 7181.9 & LT & -- & 21.97$\pm$0.27 & -- & -- & -- & -- & 21.11$\pm$0.11 & 19.63$\pm$0.06 & 18.74$\pm$0.03 & 18.07$\pm$0.06&\\
 114.0 & 7184.0 & LT & -- & 22.48$\pm$0.51 & -- & -- & -- & -- & 21.40$\pm$0.15 & 19.67$\pm$0.04 & 18.73$\pm$0.06 & 18.10$\pm$0.04&\\
 132.1 & 7202.1 & LT & -- & -- & -- & -- & -- & -- & 22.12$\pm$0.36 & 20.19$\pm$0.08 & 19.06$\pm$0.06 & 18.28$\pm$0.04&\\
 136.1 & 7206.1 & NOT & -- & -- & 21.00$\pm$0.23 & 19.73$\pm$0.08 & -- & -- & -- & -- & -- & --&\\
 197.9 & 7267.9 & NOT & -- & -- & -- & 22.41$\pm$0.46 & 19.98$\pm$0.19 & -- & -- & -- & -- & --&\\
 199.9 & 7269.9 & LT & -- & -- & -- & -- & -- & -- & -- & 22.81$\pm$0.50 & 20.81$\pm$0.14 & --&\\
 202.9 & 7272.9 & LT & -- & -- & -- & -- & -- & -- & -- & 23.45$\pm$0.58 & 21.00$\pm$0.17 & --&\\
 210.9 & 7280.9 & LT & -- & -- & -- & -- & -- & -- & -- & -- & 21.74$\pm$0.27 & 19.68$\pm$0.05&\\
 303.0 & 7373.0 & SAO$-$6m & -- & -- & -- & $>$22.98 & $>$21.49 & -- & -- & -- & -- & --& [2]\\
 382.0 & 7452.0 & Keck & -- & -- & -- & -- & $>$24.99 & -- & -- & -- & -- & --&\\
\enddata 
\tablecomments{References: [1] \cite{ATel7206}, 
[2] \cite{ATel8599}. } 
\end{deluxetable*}

\clearpage

\twocolumngrid
\begin{deluxetable}{rccccc} 
\tabletypesize{\footnotesize} 
\tablecolumns{6} 
\tablecaption{ PTF follow-up data of M101-OT.\label{table:ptf}} 
\tablehead{ 
\colhead{ Phase}  & \colhead{ MJD } & \colhead{ Telescope } & 
 \colhead{ $m_g$ }  & 
\colhead{ $m_r$ }  & \colhead{ $m_i$ }  \\ 
 \colhead{ (days) } 
& \colhead{  (+50000)    } 
& \colhead{   } & \colhead{ (mag)} & \colhead{ (mag) } & \colhead{ (mag) } \\ 
} 
\startdata 
\input{ptf}
\enddata 
\end{deluxetable}

\onecolumngrid

 \begin{table*}
 \begin{small}
 \caption{NIR and MIR photometry of M101-OT.}
 \centering
 \begin{tabular}   {rp{1.2cm}cccccc}
 \hline
    Phase & MJD     & Telescope& $J$ & $H$ & $K$ & $3.6\mu m$ & $4.5\mu m$ \\ 
    & (+50000) & & (mag) &  (mag)& (mag) & (mag) & (mag)\\ \hline
  \hline
  -4065.0 & 3005.0 & Spitzer & -- & -- & --& $>$19.00& $>$18.60\\
  -3997.5 & 3072.5 & Spitzer & -- & -- & -- & --& $>$20.48\\
  -2075.0 & 4995.0 & CFHT& 22.74$\pm$0.53 & --& 19.54$\pm$0.11 & -- & --\\
  -1756.6 & 5313.4 & UKIRT& 20.61$\pm$0.16 & -- & -- & -- & --\\
  17.1 & 7087.1 & NOT& 15.48$\pm$0.03& 15.10$\pm$0.06& 14.97$\pm$0.09 & -- & --\\
  48.0 & 7118.0 & NOT& 16.42$\pm$0.03& 15.74$\pm$0.05& 15.47$\pm$0.01 & -- & --\\
  62.6 & 7132.6 & Spitzer & -- & -- & --& 14.72$\pm$0.01& 14.75$\pm$0.01\\
  74.0 & 7144.0 & NOT& 16.70$\pm$0.04& 15.72$\pm$0.02& 15.49$\pm$0.07 & -- & --\\
  90.3 & 7160.3 & Spitzer & -- & -- & --& 14.55$\pm$0.01& 14.59$\pm$0.01\\
  102.0 & 7172.0 & NOT& 16.63$\pm$0.03& 15.90$\pm$0.04& 15.41$\pm$0.06 & -- & --\\
  121.8 & 7191.8 & Spitzer & -- & -- & --& 14.50$\pm$0.01 & --\\
  136.0 & 7206.0 & NOT& 16.53$\pm$0.02& 16.01$\pm$0.03& 15.46$\pm$0.04 & -- & --\\
  150.8 & 7220.8 & Spitzer & -- & -- & --& 14.39$\pm$0.01 & --\\
  197.9 & 7267.9 & NOT& 17.16$\pm$0.03& 16.02$\pm$0.03& 15.25$\pm$0.02 & -- & --\\
  202.7 & 7272.7 & Spitzer & -- & -- & --& 14.30$\pm$0.01& 14.50$\pm$0.01\\
  253.3 & 7323.3 & NOT & --& 17.87$\pm$0.07& 16.32$\pm$0.03 & -- & --\\
  280.3 & 7350.3 & NOT & --& $>$16.55& 16.81$\pm$0.07 & -- & --\\
  339.0 & 7409.0 & SAI-2.5m& 18.65$\pm$0.05& 17.64$\pm$0.10& 16.32$\pm$0.05 & -- & --\\
  385.0 & 7455.0 & SAI-2.5m& 21.00$\pm$0.20& 20.20$\pm$0.30& 18.20$\pm$0.20 & -- & --\\
  403.4 & 7473.4 & P200& 21.84$\pm$0.42& 19.59$\pm$0.17& 17.90$\pm$0.12 & -- & --\\
 438.0 & 7508.0 & NOT & -- & --& 17.22$\pm$0.12 & -- & --\\ 
  \hline
 \end{tabular}
 \label{table:irphot}
 \end{small}
 \end{table*}

\label{lastpage}

\bibliographystyle{aasjournal}
\bibliography{m101_bib}